# The dominant mechanisms for the formation of solute-rich clusters in low-Cu steels under irradiation.

N. Castin[(1,*)], G. Bonny[(1)], A. Bakaev[(1)], F. Bergner[(2)], C. Domain[(3)], J.M. Hyde[(4,5)], L. Messina[(6)], B. Radiguet[(7)] and L. Malerba[(1,8,*)]

[(1)] Studiecentrum voor Kernenergie – Centre d'Études de l'énergie Nucléaire (SCK•CEN), NMS Unit, Boeretang 200, B2400 Mol, Belgium.
[(2)] Helmholtz-Zentrum Dresden-Rossendorf, Institute of Resource Ecology, P.O. Box 510119, 01314 Dresden, Germany.
[(3)] EDF-R&D, Département Matériaux et Mécanique des Composants (MMC), Les Renardières, F-77818 Moret sur Loing Cedex, France.
[(4)] National Nuclear Laboratory, Culham Science Centre, Abingdon, Oxfordshire OX14 3DB, UK.
[(5)] Department of Materials, University of Oxford, Parks Road, Oxford OX1 3PH, UK
[(6)] KTH Royal Institute of Technology, Nuclear Engineering, 106 91 Stockholm, Sweden.
[(7)] Groupe de Physique des Matériaux, Université et INSA de Rouen, UMR CNRS 6634, B.P. 12, 76801 Saint-Etienne du Rouvray Cedex, France.
[(8)] Centro de Investigaciones Energéticas, Medioambientales y Tecnológicas (CIEMAT), Avda. Complutense 40, 28040 Madrid, Spain.

[(*)] Corresponding author: nicolas.m.b.castin@gmail.com

## Abstract

The formation of nano-sized, coherent, solute-rich clusters (NSRC) is known to be an important factor causing the degradation of the macroscopic properties of steels under irradiation. The mechanisms driving their formation are still debated. This work focuses on low-Cu reactor pressure vessel (RPV) steels, where solute species are generally not expected to precipitate. We rationalize the processes that take place at the nanometre scale under irradiation, relying on the latest theoretical and experimental evidence on atomic-level diffusion and transport processes. These are compiled in a new model, based on the object kinetic Monte Carlo (OKMC) technique. We evaluate the relevance of the underlying physical assumptions by applying the model to a large variety of irradiation experiments. Our model predictions are compared with new experimental data obtained with atom probe tomography and small angle neutron scattering, complemented with information from the literature. The results of this study reveal that the role of immobilized self-interstitial atoms (SIA) loops dominates the nucleation process of NSRC.

# 1 Introduction

Assessing and predicting embrittlement of reactor pressure vessel (RPV) steels as a function of the neutron dose is of crucial importance for the safe operation of nuclear power plants. Substantial effort has been put over the last decades not only to develop so-called dose-damage correlations of semi-empirical nature [1], but also to ensure they are based on mechanistic understanding [2]. The data to which these correlations are fit come from vessel surveillance [3], decommissioned vessels [4], or irradiation campaigns in materials testing reactors (MTR) [5-7]. For a longer-term application, effort has been put to develop fully physically informed suites of computer simulation codes, aimed at predicting RPV steel radiation hardening [8,9], including microstructural examination studies in support of modelling [10-28]. These studies revealed that radiation hardening in RPV steels, as well as in other types of iron alloys such as high-Cr ferritic-martensitic (F-M) materials, is mainly the consequence of the formation of high densities ($\sim 10^{23}$ $m^{-3}$) of nanometre-size solute-rich clusters (NSRC), which act as obstacles to dislocation motion [29,30]. Importantly, NSRC are also found in steels containing low quantities of Cu for which all solute species are found below their solubility limit in the binaries.

Embrittlement in RPV steels ensues mainly from hardening, due to the fact that the obstruction of dislocation motion hinders the blunting of crack tips. Contributions from embrittlement without hardening cannot be excluded, due to solute segregation at grain boundaries that can result in intergranular fracture [31], but these are generally expected to have a minor role in determining brittle behaviour.

Microstructure examination experiments show that, after irradiation, the NSRC contain varying concentrations of Cu, Mn, Ni, Si and P, as well as Cr in the case of F-M steels. Due to their small size and coherence with the matrix (body centred cubic – bcc – structure), they are not easily resolvable to conventional transmission electron microscopy (TEM). Their existence has been revealed mainly through other microstructural characterization techniques, such as atom probe tomography (APT) [10-21,32] and small angle neutron scattering (SANS) [24-28]. Recently, though, modern scanning TEM (STEM) techniques coupled with energy dispersive spectroscopy (EDS) allowed these clusters to be resolved, as well [33].

Empirical correlations reveal a direct proportionality between hardening ($\Delta\sigma_y$, where $\sigma_y$ is the yield strength and $\Delta$ indicates the increase) or embrittlement ($\Delta T_{NDT}$, where $T_{NDT}$ is the *nil ductility temperature*), and the square root of the volume fraction of these clusters [28,34]. Physical models suggest the dependence of hardening on the square root of obstacle density times their size [35,36]. Therefore, a quantitative assessment of radiation-induced hardening and embrittlement as a function of radiation dose and dose-rate requires a correct description of the kinetics of the atomic-level processes that lead to the formation of the NSRC observed by APT, SANS, STEM-EDS, etc., which act as obstacles to dislocation motion. The mentioned experimental techniques provide invaluable information concerning number density, size and composition of NSRC after irradiation experiments. However, only physical models can describe the atomic-level processes that lead to their formation in sufficient detail, allowing the correct assessment of these quantities also without the need of microstructural examinations that are not routinely performed on materials in operation, for any dose and at any dose-rate. In particular, the knowledge of the kinetics of formation is essential with a view to predicting correctly the build-up of damage versus dose, so as to be able to anticipate, as much as possible, the materials degradation up to doses of relevance for, e.g., long term operation.
The involved processes stem from irradiation, which introduces in crystalline materials an excess of point defects (vacancies and self-interstitial atoms) that are produced in cascades of atomic displacements (displacement cascades). These are complex, collective processes that are triggered by the collision of the incoming energetic particles (neutrons in nuclear reactors) with the atoms of the lattice and have been studied for decades by means of atomistic simulation tools [37-39].

Because of the complexity of the physical processes induced by the continuous injection of point defects during irradiation, the mechanisms leading to the subsequent formation of NSRC are still debated. Previous work emphasized the role of driving forces from thermodynamics [40-45]. In this framework the role of radiation defects (particularly vacancies) would be to accelerate the diffusion processes that lead the system to thermodynamic equilibrium in terms of phase separation, because their concentration is well above the equilibrium one. In contrast, other researchers have suggested that the evolution of the system is dominantly driven by mechanisms specific to irradiation [16, 21, 46, 47], i.e. considering as crucial the role of self-interstitial atoms (SIA). This work aims at providing a quantitative estimate of the dominant mechanisms that govern the formation of NSRC, considering both the role of vacancies and SIAs, in the light of the latest theoretical and experimental evidence concerning atomic-level processes of diffusion and transport. For this reason, i.e. to collect and juxtapose all the relevant information, section 2 is devoted to reviewing the work that has been done recently on these subjects. New experimental evidence from APT and SANS obtained in this work is reported, in complement to information from the literature. This review is instrumental to be able to draw a detailed and complete map of the mechanisms that can be considered responsible for the nucleation and growth of the NSRC. Next, in section 3, these mechanisms are incorporated in a new model that describes the evolution of the radiation-induced nano-features in ferritic steels, by implementing for the first time explicit solute transport in an object kinetic Monte Carlo framework. The formation of NSRC is one of the natural outputs of this model, which encompasses all features related with irradiation effects at the nanometre scale, importantly including point-defects clusters. The model is then applied, in section 4, to simulate irradiation experiments on a wide variety of RPV steels and model alloys. Direct comparison between the model prediction and experimental evidence with APT enables the relevance of the physical assumptions that underlie the model to be tested. Our findings are then discussed in section 5, where we deduce the dominant mechanism at play.

# 2 Mechanism driving the formation of nano-sized solute rich clusters

In this section we explain in detail our model, showing that it is rooted in extensive theoretical and experimental studies performed over the last decade.

## 2.1 Theoretical considerations

Calculations with density functional theory (DFT) reveal that all the solute atoms that are typically observed in NSRC (Cu, Mn, Ni, Si and P) exhibit attractive binding energies with vacancies [48, 49]. This has two consequences:

1. These solutes are transported by single vacancies via a drag mechanism. To be more precise, the flux of vacancies is coupled to the flux of these solutes and oriented in the same direction (positive correlation). This naturally leads to segregation of these atoms at sinks where vacancies are absorbed. Messina *et al.* [49] demonstrated that, contrary to widespread perception, this dragging is a general process, common to many solute species in Fe. The dragging is due to the fact that the presence of the solute modifies locally the energy barriers for vacancy migration. The model by Messina *et al.* was based on an exact analytical self-consistent mean-field approach, making use of DFT-calculated migration barriers. The conclusion was that, in the infinite dilution limit, the vacancy-solute flux stemming from this positive correlation is sufficiently strong for Cu, Ni, Mn, Si and P to produce radiation-induced segregation at point defect sinks up to relatively high temperatures. The range includes 300°C, which is the approximate operation temperature of the RPV, and in fact extends to the whole ferromagnetic temperature range. On the contrary, a negative correlation and thus an absence of drag was predicted for typical solutes that are present in RPV and also F/M steels, but are never found in NSRC, e.g. Mo.
2. Stable vacancy-solute clusters are expected to form. In this work (see later in section 3), we have evaluated the magnitude of the collective binding energies between solutes and vacancies by performing a large amount of DFT calculations, up to triplet configurations (see supplementary material for the full tabulation). The results of these calculations demonstrate that solute-vacancy pairs can remain stable defects at RPV-relevant temperatures (near 300°C). We thus expect these configurations to be nuclei for the formation of larger clusters.

DFT also reveals a significant interaction between solute atoms and SIA defects, namely:

3. Some solutes, specifically P, Mn and Cr, are transported by SIAs, via the formation and migration of mixed dumbbells [50]. By definition, this implies the migration of the involved point-defects and solutes in the same direction. Hence, once again, solutes are transported towards point-defect sinks. It has been calculated that these three solutes are indeed prone to segregation at point defect sinks, due to SIA transport, up to high temperatures [48].
4. It has also been shown that, in addition to small SIA clusters [51], small prismatic loops of self-interstitial nature that may form directly in displacement cascades interact strongly and attractively with solutes, especially with P, Si, Mn, Cu and Ni (in order of strength, from the strongest to the weakest interaction) [52]. The interaction energy depends on whether the solute interacts with the centre or the edge of the loop; it ranges between 0.2 and 0.5 eV and can be as high as 1 eV in the case of P. In addition, both experiments and atomistic studies have shown that carbon-vacancy complexes (possibly nitrogen-vacancy and oxygen-vacancy complexes as well [53]) form abundantly under irradiation [54,55] and act as very efficient traps for gliding prismatic loops [56], with interaction energies as high as 1.3 eV [57]. The existence of an affinity between solute atoms and loops, as revealed by DFT, has been extended to atomistic studies with interatomic potentials using Metropolis Monte Carlo techniques. These studies revealed that Cu, Mn, Ni, and P (Si has not been studied because of the absence of a suitable Fe-Si interatomic potential) naturally tend to accumulate around extended defects, such as dislocation loops and lines. They thus create energetically stable clouds around them, as well as precipitates (the simulations specifically suggest the formation of Cu and NiMn B2 phase precipitates, or Mn-rich regions, attached to dislocation loops and lines) [58,59].

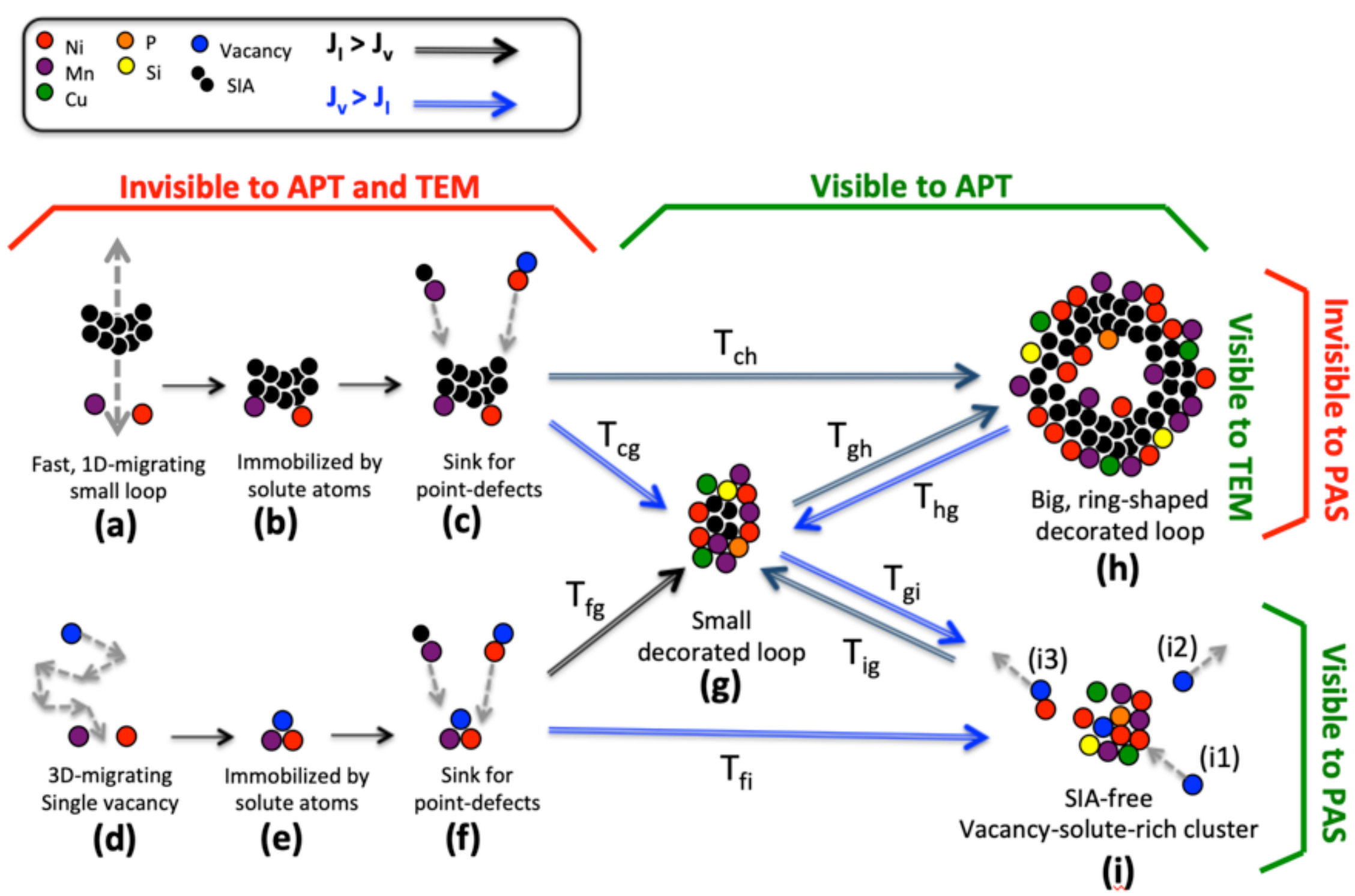

**Fig. 1** – Hypothetical scenario for the formation of NSRC. Solute atoms and point-defects are represented by coloured circles (see legend). Fe is the matrix, not depicted in the figure. $J_I$ and $J_V$ denote the incoming flux in interstitial defect or vacancy defects, respectively.

Based on these results of previous work, we propose here the general scenario that is illustrated in Fig. 1 for the nucleation of NSRC in steels under irradiation. A first sketch of this possible scenario has been put forward for FeMnNi alloys [60-61], but it is here generalised and fully developed. To start with, two distinct mechanisms of NSRC nucleation may exist:

1) In Fig. 1a-1c, nucleation occurs because of the interaction between SIA-defects and solute atoms.
   a. In Fig. 1a, small, glissile, one-dimensionally migrating SIA clusters (i.e. small prismatic dislocation loops) are created in the material during or after displacement cascades [37-39, 62]. Their migration energy is ~0.1 eV or less in the absence of interaction with impurities [63].
   b. In Fig. 1b, the migrating SIA cluster encounters solute atoms (or carbon-vacancy complexes) along its glide prism. Its migration is hence temporarily stopped, due to the attractive interaction, the strength of which depends on the nature of the solute (or complex) and its relative position with respect to the loop (centre or edge) [52]. While in most cases a relatively weak binding energy is involved, some solutes, like P, or pairs of solutes, as well as carbon-vacancy complexes, can trap the loop with energy values around, or in excess of, 1 eV. At the RPV operation temperature, this energy is sufficient to immobilise the defect completely. Clearly, these types of encounters are more likely with increasing solute concentration. DFT studies [52] suggest, in addition, that the interaction with multiple solutes adds up in strength, thereby making weakly interacting solutes like Cu, Ni and Mn also able to act as traps, when combined. Since a single prismatic dislocation loop cannot, by gliding, drag substitutional solutes with it, this interaction does not lead to net solute transport [64].

c. In Fig. 1c, provided that the time of immobilization is long enough, the loop effectively becomes a sink for migrating point defects. Since many point-defects are expected to carry solute atoms, there is a high chance that new solutes are thereby deposited on the loop. The loop remains thus trapped with increasing strength, eventually losing any possibility of migrating away from the trapping solutes, while the local concentration of solutes keeps increasing. The latter effect, combined with the strong affinity that some solutes have for loops [52], makes these locations suitable also for the precipitation of equilibrium thermodynamic phases, including some that would not precipitate, if the concentration remained equal to the nominal one, but will precipitate because locally the concentration exceeds the solubility limit [65]. Importantly, this process favours the accumulation of *solutes irrespective of whether they do or do not form a known stable thermodynamic phase*. Thus, the composition of NSRCs may, but need not, correspond to the composition of stable thermodynamic phases. In addition, recent thermodynamic studies [66] based on DFT revealed that the number of phases that may exist in RPV-steel like systems is extremely large and some phase competition could occur, well beyond the number of phases included e.g. in Thermo-Calc [67]. Thus, the variety of potentially possible compositions is wide.

2) In Fig. 1d-1f, a similar scheme to the one above is anticipated, this time pivoting around the interaction between vacancy defects and solute atoms:

a. Like SIA defects, migrating vacancies, or migrating small vacancy-clusters in Fig. 1d, are likely to encounter solute atoms. Mutual binding between vacancies and solutes leads to the formation of complexes, as revealed by our DFT study. Small solute-vacancy clusters can therefore form, as depicted in Fig. 1e.

b. Consequently, in Fig. 1f these vacancy-solute clusters are also expected to play the role of sinks for other migrating point-defects. As the latter carry solute atoms, the local enrichment increases (similarly to Fig. 1c). It is important to note that, in contrast to SIA-solute clusters in Fig. 1c, solute-vacancy clusters can dissolve at RPV-relevant temperatures. Since the magnitude of the binding energy for a single vacancy, or a vacancy-solute pair is of the order of 0.3-0.7 eV, the emission of single defects from the cluster constantly occurs, also provoking the dissolution of solute species.

Following the nucleation events described above, Fig. 1 also depicts possible growth mechanisms, via transitions indicated by thick arrows. Since loops cannot transport solutes, and assuming that the transport of solutes clusters by vacancies is negligible, no direct coalescence of NSRC is expected. Instead, immobilized defects in Fig. 1c and Fig 1f are affected by the incoming flux of mobile single point defects, which is always associated with a positive flux of solute atoms. Given a particular defect, we define as $J_I$ the incoming flux of self-interstitial atoms towards it, and as $J_v$ the incoming flux of vacancies towards it:

c. The case $J_I > J_v$, i.e. if the incoming flux is predominantly of SIA kind (including single SIAs but also small SIA clusters not yet immobilized by solute atoms), thus outnumbering the incoming flux of vacancies, is depicted by thick black arrows in Fig. 1. A vacancy-solute cluster in Fig. 1f would thus follow the transition $T_{fg}$, gradually losing its vacancy content, until becoming a small decorated loop like in Fig. 1g. Should the magnitude of $J_I$ be large enough, the defect can even follow the $T_{gh}$ transition and become a large loop like in Fig. (1h). Since the binding between solutes and loops is stronger at the periphery of the latter [52], solutes are expected to accumulate there, giving rise to the ring shape depicted in the figure. Defects originally nucleated as in Fig. 1a-1c also follow the same fate.

d. The case $J_v > J_I$, i.e. if the net balance of incoming point defects is higher for vacancies than SIA, is depicted by thick blue arrows in Fig. 1. Small loops like those in Fig. 1g are in this case expected to shrink, eventually becoming SIA-free defects, as depicted in Fig. 1i. Defects originally nucleated as in Fig. 1d-1f also evolve towards this configuration.

Our model in Fig. 1 therefore foresees the existence of three distinct kinds of solute-rich clusters. They are all visible using APT, provided that the total number of solute atoms involved is large enough.

- In Fig 1g, small loops that may not be visible to the TEM because of the too small number of SIA defects are, however, visible to the APT, thanks to their solute decoration. They would be seen by this experimental technique as a spherical cloud of solute atoms, while the SIA content cannot be resolved. It is important to note that the interstitial nature of the loop makes it very unlikely that a SIA is released from it, because the binding energy of SIAs to mother clusters is on the order of several eV, except for clusters of 2-3 SIA [68]. On the other hand, vacancies approaching the defect will recombine and therefore will not produce any solute transport or rearrangement. Consequently, it is very unlikely that solutes can be removed from a decorated self-interstitial loop.
- In Fig. 1h, big solute-decorated loops are foreseen as a possible limiting case from Fig 1g, given favourable conditions. These defects are visible to the TEM, because of their loop nature. Their typical ring-shape is likely to be visible to the APT as well, if homogeneously decorated by solute atoms. However, solute atoms may also accumulate inhomogeneously, giving rise to groups of separate solute clusters lying on the habit plane of the loop.
- In Fig. 1i, the solute-rich cluster is free of SIA. It has a vacancy content instead, which can be single vacancies, or small vacancy clusters. It is thus not visible to the TEM, but may be seen as a diffuse cloud of atoms by APT; moreover, positron annihilation spectroscopy (PAS) will be sensitive to the presence of these clusters and may help detect them. The solute cluster may grow further (positive incoming flux of solute atoms), but may also dissolve. Indeed, as illustrated in Fig. 1i, here the release of vacancies (Fig. 1i2) and thus also the emission of solute-vacancy pairs (Fig. 1i3), is possible, so these types of clusters are thermally less stable. This reminds of the so-called "unstable matrix damage" sometimes invoked to explain neutron dose rate effects in RPV steel embrittlement [69,70].

In summary, the conceptual model depicted in Fig. 1 and based on the most relevant theoretical evidence anticipates that NSRC nucleation can take place from the mutual interaction between solute atoms with vacancies and SIA defects. Because of the strong affinities between point-defects and solute atoms, 'nuclei' for NSRC can be as small as one or two solutes. The catalysing effect of point defect clusters requires no solute-cluster critical size that should be overcome so that the interface energy is compensated by the gain in free energy due to phase separation, as would happen in a "classical" nucleation-and-growth scenario.

## 2.2 Experimental evidence

Many pieces of experimental evidence support the conceptual model described above in section 2.1 for NSRC formation, as depicted in Fig. 1.

To start with, APT frequently reveals segregation of solute atoms on various crystal defects such as dislocation lines, grain boundaries or carbide matrix interfaces in irradiated RPV steels [16-23, 32]. This clearly supports the hypothesis that solute atoms are continuously dragged by single point defects towards point-defect sinks.

Next, a large number density (~$10^{22}$ $m^{-3}$ to $10^{23}$ $m^{-3}$) of roughly spherical NSRC is systematically reported in nearly all APT studies on irradiated steels [10-24]. A typical image obtained from APT is shown in Fig. 2. These are clearly associated with the defects depicted in Fig. 1g and Fig 1i, as discussed above. A collection of experimental data is summarized in Tab. 1, for a total of 10 materials, that are divided in three categories: (1) model alloys for RPV steels; (2) base metals in RPV steels; (3) weld metals in RPV steels. Irradiation conditions in terms of dose rate and temperatures also differ through Tab. 1, as sample materials were irradiated either in operating nuclear power plants, or in MTR facilities. The majority of these data were obtained during the LongLife European project [71], but had remained so far unpublished (they are therefore denoted as "this work" in Tab. 1).

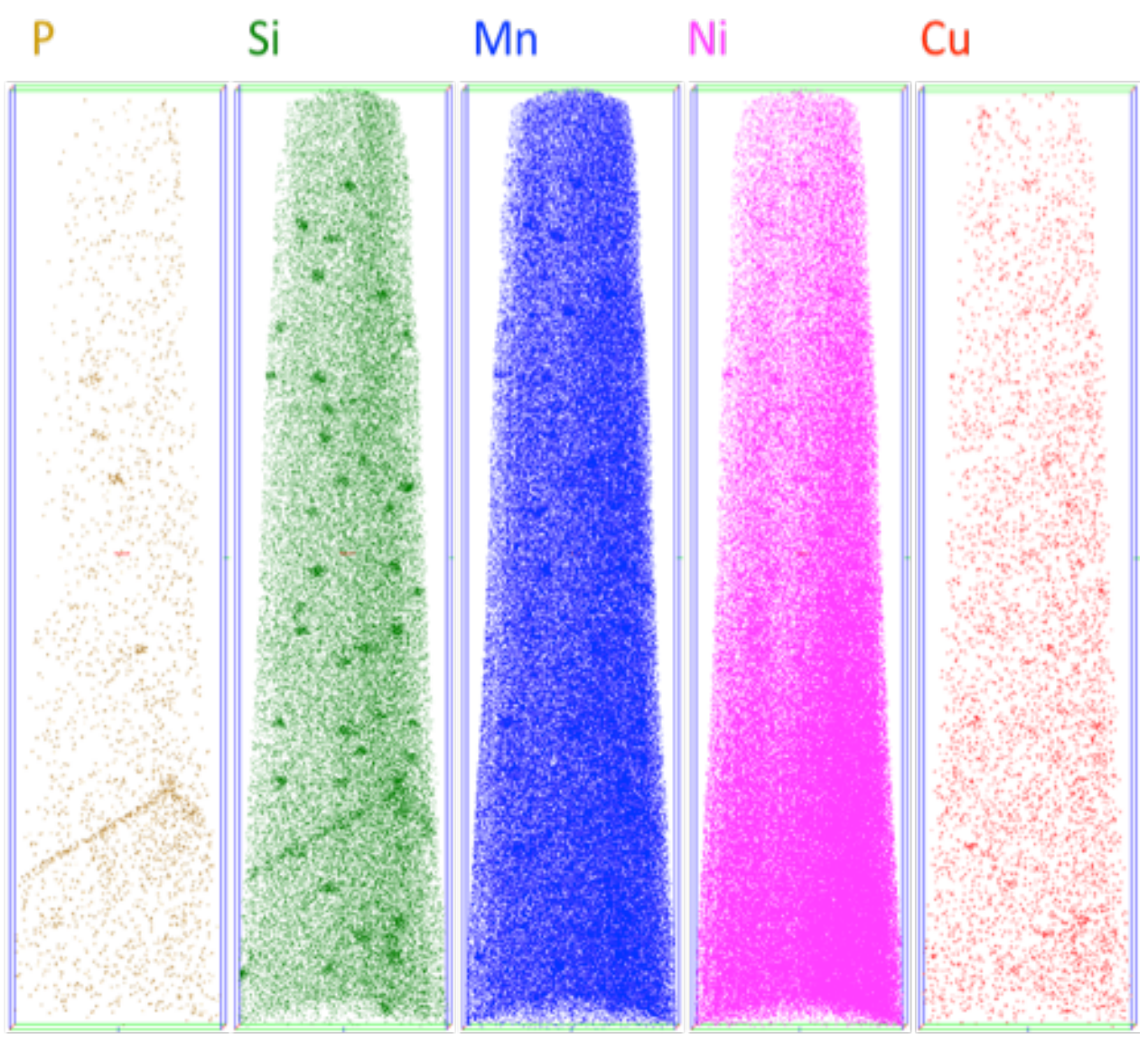

**Fig 2 –** Atom maps showing the distribution of solute atoms in a small volume of a typical French RPV steel (base metal) irradiated with neutrons at 290°C [14]. The sample has a volume of 45×45×190 $nm^3$.

Studies in model alloys for RPV steels using PAS techniques suggest the existence of complexes with up to 10 vacancies associated with solutes [72-75]. This proves that the clusters depicted in Fig. 1i as a possible consequence of the assumptions of our model actually exist. Interestingly, PAS studies also reveal that the majority of vacancy-type defects are in fact found in a Fe-rich environment, i.e., they are not associated with a local high solute concentration. This can be appreciated by looking at S-W curves, e.g. in Fig. 8 in Ref. [76]. It is seen that these curves, for irradiated RPV steels, are close to the one corresponding to pure Fe, and not to those associated with pure Ni or pure Mn materials. This indirectly suggests the dilute nature of the clusters against the compact shape that would be expected for nuclei of specific thermodynamic phases.

Fig. 3 summarizes the characteristics of the NSRC as seen by APT, for all cases listed in Tab. 1. It reveals no clear trends between density/size of NSRC and radiation dose, except for a weak tendency to slowly increase the number density and the volume fraction with dose, as is substantiated by a power law fit. The number density is generally measured to lie between $10^{22}$ m$^{-3}$ and $10^{23}$ m$^{-3}$ in the whole range of doses from 0.01 to almost 1 dpa. The mean size of the NSRC lies between 1 and 4 nm, with an average around 2.5 nm and essentially no increase with dose. Importantly, the volume fraction exhibits no sudden appreciable increase above a certain incubation dose. This suggests that the mechanism that is responsible for the formation of NSRC deviates from a classical nucleation and growth process: solute clusters keep more or less constant size, as if solutes were equally distributed in each of them, and slowly grow in number.

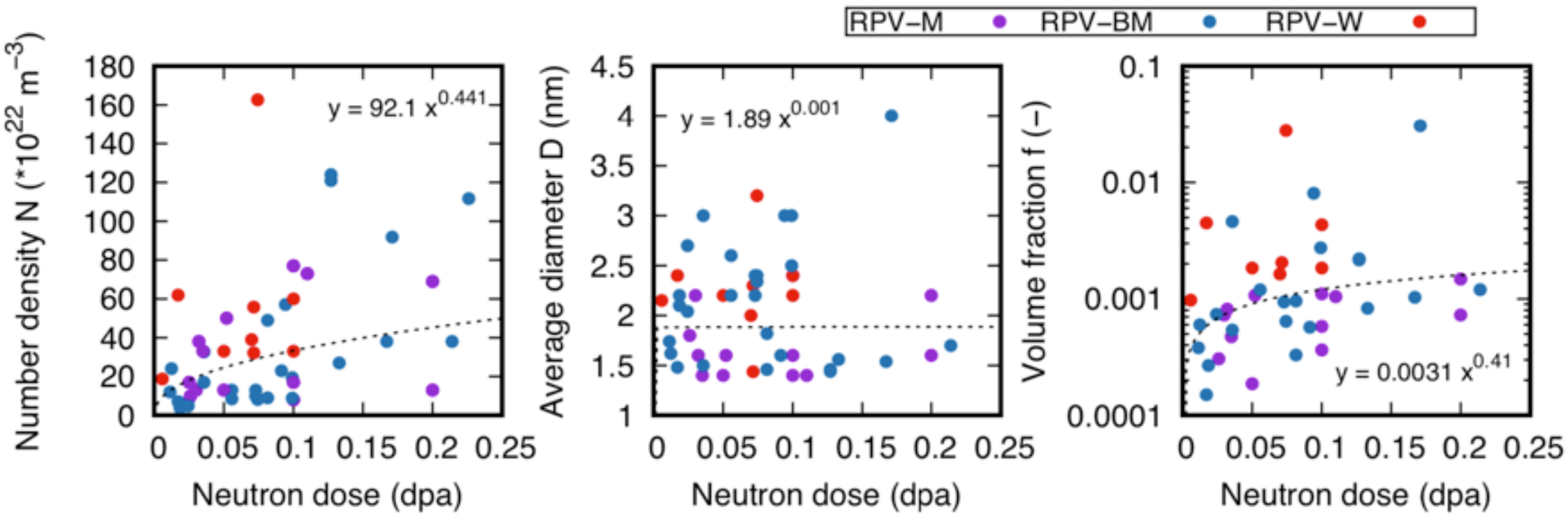

**Fig. 3** – Evolution with the received neutron dose of the number density *N*, average size *D* and volume fraction ($f = ND^3\pi/6$) of the NSRC, as evidenced by APT and SANS, and used as reference data in this work (see description and references in Tab. 1). Dashed lines show power law regressions.

Finally, experimental evidence for defects as depicted in Fig. 1h is found in the literature. The point-defect content of NSRC cannot be deduced by APT. Nevertheless, if the loop is large enough, due to the particularly favourable edge location, rings of solutes have been detected, for example toroidal segregations of Mn, Ni and Si were observed in Russian VVER base metals [71]. Coincidentally, dislocation loops were observed by TEM in the same materials [77], with densities (~$10^{22}$ m$^{-3}$) and sizes (about 5 nm) compatible with these toroidal APT solute clusters (respectively 6-7nm and $10^{22}$ m$^{-3}$). Similar APT observations of solute rings have been reported in [78]. Alternatively, lumps of solute clusters decorating discontinuously the loop edge line have been detected after careful analysis of APT results, as a system of close clusters lying on the same atomic plane [79]. Planar clusters of solute atoms were reported also in APT studies on VVER materials [13], as well as in austenitic steels (although in the latter case at a significantly higher dose, i.e., 10 to 28 dpa) [78]. These are all clear evidences that solute clusters are indeed associated with large loops in some irradiated steels, as stemming from the assumption of the model (Fig. 1h). Such clusters are, however, far less common than spherical clusters. To the authors' knowledge, they have never been observed in Western RPV materials, in the range of chemical compositions and irradiation conditions that are listed in Tab. 1. This is consistent with the physics of irradiation-induced defects, considering the differences of diffusion properties between SIA and vacancy defects: SIA defects tend to be one-dimensionally migrating, thus the sink strength towards sinks diluted in the bulk of the materials is lower. Consequently, in Fig. 1, $J_V$ is, in a majority of cases, higher than $J_I$.

**Table 1 –** List of chemical composition and irradiation conditions for the RPV materials investigated in this study. The last column indicates the range of numerical values for the $E_{Sat}^{(bind)}$ parameter that maximizes the agreement between OKMC simulations and experimental data. The abbreviations LDR, MDR and HDR stand for "low/medium/high neutron dose rate", respectively. Data denoted as "this work" were obtained during the LONGLIFE European Project [71].

| Name | Reactor | Temp (°C) | Dose rate (dpa/s) | Grain size (μm) | Disl. Dens. ($10^{14}$ m$^{-2}$) | C (appm) | Cu (at%) | Ni (at%) | Mn (at%) | Si (at%) | P (at%) | Cr (at%) | Exp. technique | Ref | $E_{Sat}^{(bind)}$ (eV) |
|---|---|---|---|---|---|---|---|---|---|---|---|---|---|---|---|
| **Model alloys for RPV steels** | | | | | | | | | | | | | | | |
| FeNiMn | MTR | 290 | $2.5\times10^{-9}$ (MDR)<br>$1.3\times10^{-7}$ (HDR) | 75 | 0.5 | 50 | - | 0.73 | 1.12 | - | 0.0045 | - | APT | [80] | 0.7 – 0.85 |
| FeCuNiMn | MTR | 290 | $2.5\times10^{-9}$ (MDR)<br>$1.3\times10^{-7}$ (HDR) | 75 | 0.5 | 50 | 0.068 | 0.57 | 0.98 | - | 0.0034 | - | APT | [10] | 0.7 – 0.8 |
| **Base metal in RPV steel** | | | | | | | | | | | | | | | |
| EDF 1 | NPP | 270 | $2\times10^{-10}$ (LDR) | 1 | 2.5 | 150 | 0.078 | 0.61 | 1.02 | 0.61 | 0.018 | 0.16 | APT | This work | 0.85 – 0.9 |
| EDF 2 | NPP | 286 | $2.5\times10^{-10}$ (LDR) | 1 | 2.5 | 150 | 0.031 | 0.51 | 1.1 | 0.36 | 0.011 | 0.21 | APT SANS | This work | 0.75 – 0.8 |
| EDF 3 | NPP | 286 | $2\times10^{-10}$ (LDR) | 1 | 2.5 | 150 | 0.029 | 0.38 | 0.86 | 1.0 | 0.016 | 0.021 | APT SANS | This work | 0.75 – 0.8 |
| FZD | NPP | 255 | $1.7\times10^{-10}$ (LDR)<br>$4\times10^{-9}$ (MDR) | 1 | 2.5 | 150 | 0.024 | 0.7 | 0.56 | 0.47 | 0.036 | 0.18 | SANS | This work | 0.7 – 0.9 |
| | MTR | 290 | $5\times10^{-9}$ (MDR)<br>$10^{-7}$ (HDR) | | | | | | | | | | | | |
| **Weld metals in RPV steels** | | | | | | | | | | | | | | | |
| Ringhals E | NPP | 275 | $2.5\times10^{-10}$ (LDR) | 1 | 2.5 | 150 | 0.05 | 1.5 | 1.25 | 0.41 | 0.01 | 0.07 | APT | [22] | 0.7 – 0.75 |
| Ringhals N | NPP | 275 | $2.5\times10^{-10}$ (LDR) | 1 | 2.5 | 150 | 0.06 | 1.97 | 1.1 | 0.28 | 0.02 | 0.07 | APT | [22] | 0.7 – 0.75 |
| ANP2 | NPP | 285 | $1.8\times10^{-8}$ (MDR) | 1 | 2.5 | 150 | 0.026 | 0.96 | 0.88 | 0.24 | 0.034 | 0.11 | APT SANS | This work | 0.7 – 0.8 |
| ANP6 | NPP | 280 | $5.7\times10^{-10}$ (LDR)<br>$4\times10^{-9}$ (MDR) | 1 | 2.5 | 150 | 0.07 | 1.61 | 1.144 | 0.3 | 0.022 | 0.075 | APT | This work | undefined |

# 3 Object kinetic Monte Carlo model for NSRC formation

The relative importance of the different mechanisms summarized in Fig. 1 to drive the process of NSRC formation, can be quantified by introducing them, with suitable parameters, in a model that describes the evolution of the nanostructure. The features of this model, conceptually described in section 2, are presented in what follows. Premises for such model were already proposed in previous work by Jansson *et al.* [81] and Chiapetto *et al.* [60-61], based on the object kinetic Monte Carlo (OKMC) method [82], but limiting the scope to simple model alloys (Fe-C, Fe-MnNi). Here we generalise those models, by adding all the required reactions for the mechanisms shown in Fig. 1, so as to describe neutron irradiation in steels (section 3.1), including explicit treatment of solute transport. This generalization is made possible by the massive calculation of parameters performed using DFT, as reported in section 3.2.

## 3.1 General description of the model

In an OKMC simulation, the evolution of a system that contains diffusing species is described stochastically. These can be point defects, their clusters, or complexes involving point defects and solutes – e.g. vacancy-solute clusters – altogether denoted as 'objects'. Each object is located at given coordinates in a simulation volume with periodic boundary conditions and has an associated reaction volume, conveniently assumed to be spherical, whenever this approximation is reasonable. Large loops have an associated toroidal volume. If mobile, the object can migrate by discrete jumps of a given distance, according to pre-defined probabilities. When the reaction volume of an object overlaps with the volume of another one, depending on the nature of the two, they may recombine (e.g. a self-interstitial with a vacancy), cluster (progressively forming cavities or dislocation loops), become trapped (if an immobile complex is formed), or be absorbed by sinks. Objects can also dissociate (e.g. by emission of a point defect) or move away from a trap, again according to pre-defined probabilities. All these are *events*. The probability for migration and dissociation events is expressed in terms of Arrhenius frequencies for thermally activated processes [83], in the framework of the transition state theory:

$$\Gamma_e = \nu_e \exp\left(-\frac{E_e^{(act)}}{k_B T}\right) \tag{1}$$

Here $\nu_e$ is the attempt frequency (or pre-factor) of the event $e$, and $E_e^{(act)}$ is the corresponding activation energy, both these parameters being an input to the model; $k_B$ is Boltzmann's constant and $T$ is the irradiation temperature expressed in K. Since the activation energy for migration or emission depends on the type and the size of the objects involved, a large number of parameters is needed for a standard simulation. These parameters may be derived using data from *ab initio* calculations and/or molecular dynamics, as well as other atomistic calculations, or from experiments whenever available. In some cases, rate theory helps to extend them to larger cluster sizes, while sometimes educated guesses are unavoidable. The parameters used in this work for the properties of vacancies and self-interstitial clusters (e.g. migration and dissociation rates, capture radii) are the result of years accumulating information and addressing increasingly complex alloys, as detailed in [60-61, 81]. Only departures from those parameters, or additions, are described in what follows.
The main and crucial addition introduced here with respect to previous work [60-61, 81] concerns the explicit treatment and redistribution of solute atoms of different chemical nature during irradiation. The work of Chiapetto *et al.* [60-61] considered the Fe-C-NiMn alloy. The effect of *all* alloying elements was accounted for in a 'grey alloy' approximation, i.e. their effect was translated into a change of the value of the parameters that define the mobility and stability of point defects and their clusters, as a function of the nominal composition: the effect was thus equal everywhere in the simulation volume and no solute redistribution could take place, because solutes were physically not in the box, only their effect. Here, in contrast, the Fe-C-CuNiMnSiP system is directly addressed, including explicitly all the alloying elements of steels used in nuclear applications that are known to produce NSRC. Doing this increases the complexity of the model, because the interaction of each solute with the other solutes and with point-defects needs to be described.

The way the model works is pictorially illustrated in Fig. 4.

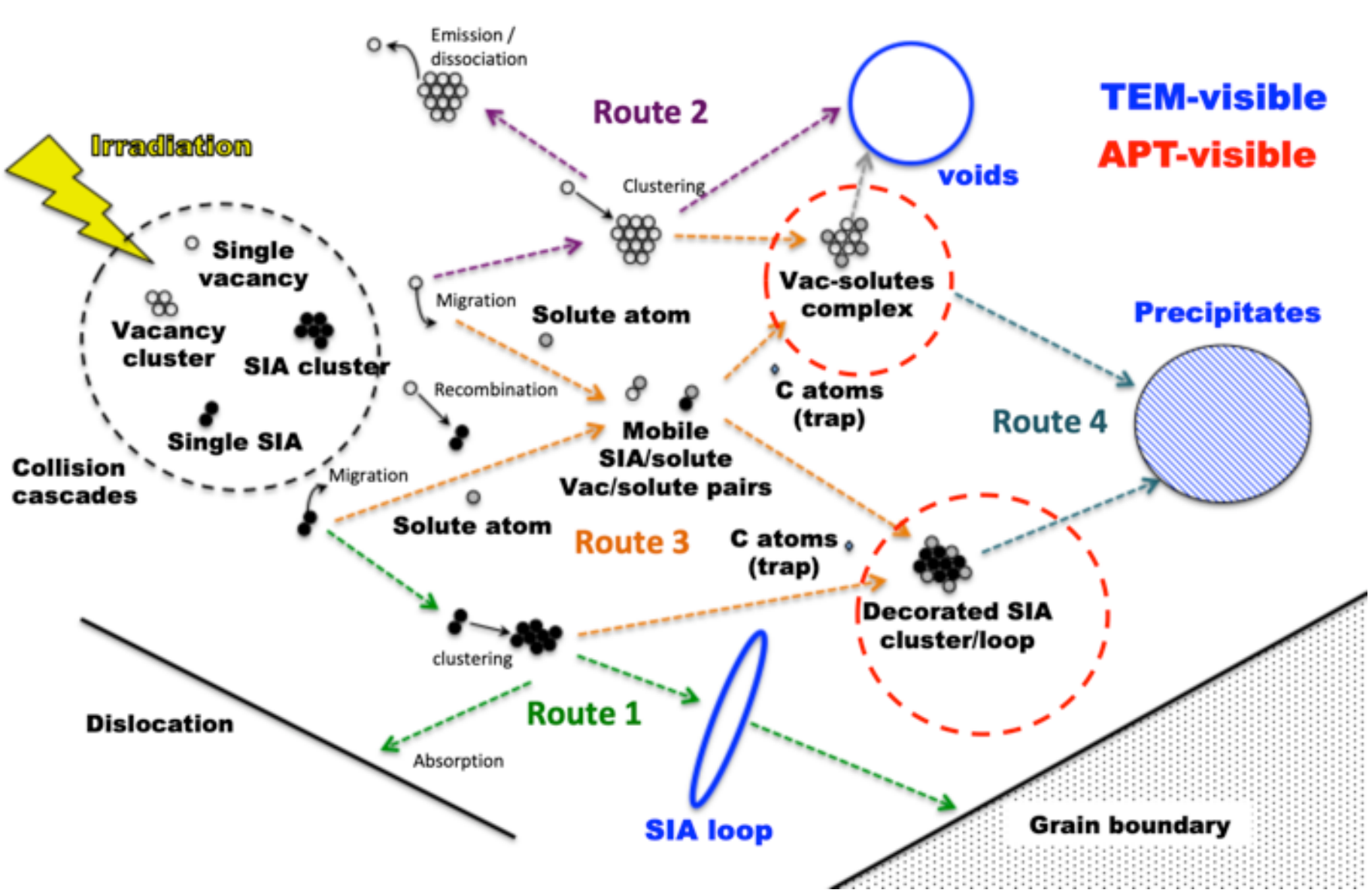

**Fig 4** – Schematic description of our OKMC model. Different routes for the injected defects are highlighted with different colours, and so are the defects visible to the TEM or the APT experimental techniques.

The initial condition is a simulation volume devoid of point defects and their clusters. Substitutional solutes, namely Cu, Ni, Mn, P, and Si, are randomly distributed in the volume, while C impurities are included as trapping objects (see detailed explanations in Refs. [60-61, 81, 84]), according to the bulk composition of the steel of interest. Due to the limited size of the simulation box, large-scale features of the microstructure are not directly introduced. They are nevertheless introduced with indirect methods aimed at accounting for their effective sink strengths:

- The dislocation network is represented by absorption events considered after each defect migration, the probability of which is calculated in order to respect the theoretical sink strength [85] due to the dislocation density observed in the material before irradiation.

- Grain boundaries are implemented following the methods proposed in past work [60-61, 81-82]. The trajectory of each object is followed in two separate sets of coordinates: (a) the coordinates in the simulation volume, where periodic boundary conditions are applied, which are used to measure distances between defects and detect mutual interactions; (b) the coordinates with respect to the centre of a spherical grain, the radius of which generally exceeds the simulation volume typical length. No periodic boundary conditions are applied in this second set of coordinates. Each newly created object is assigned a precise initial position in the first set of coordinates, according to the reactions that generated it during the OKMC simulation. On the contrary, the initial coordinates in the second set of coordinates are randomly chosen, in order to guarantee the correct sink strength by the grain boundary [85]. They are updated every time the object migrates leading to its annihilation when its position coincides with the boundary of the spherical grain.

These methods are appropriate for modelling bulk phenomena. However, processes such as segregation at grain boundaries or along dislocation lines, cannot be directly addressed if the grain size is larger than simulated volume or if less than a dislocation segment in the simulated volume.

Irradiation events, i.e. the appearance of a displacement cascade or the production of isolated Frenkel pairs (vacancy + self-interstitial), take place in a stochastic manner, based on the dose rate assigned, expressed in displacements-per-atom per second (dpa/s), applying the Norgett-Robinson-Torrens (NRT) equation [86, 87]. This choice is made to allow direct comparison with the dpa values as given in experiments. In first approximation, for all OKMC simulations in this work we used the same spectrum of energies associated with the primary knock-on atom (PKA), as derived for neutron irradiation in Ref. [88]. This can be nevertheless easily adapted to other irradiation conditions, such as with self-ions. A displacement cascade is, from the point of view of the model, a pre-defined list of point defects and their clusters, with defined positions with respect to the centre of mass of the cascade. Cascades are taken randomly from a database produced using molecular dynamics. Each is randomly rotated and injected into a randomly selected area of the simulation volume, as depicted in the left corner in Fig. 4. The thermally activated diffusion of the defects then takes place, as the Monte Carlo algorithm selects migration events. Some correlated recombination (i.e. within the cascade region) is likely to occur, but most surviving defects will diffuse to longer distances, following different possible routes (referring to Fig. 4):

- **Route 1:** SIA defects migrate in first instance because, in the absence of a significant concentration of solutes or impurities, their migration energy is low: < 0.35-0.5 eV for three-dimensionally (3D) migrating single SIAs or small clusters thereof, and as low as ~0.1 eV or less for one-dimensionally (1D) gliding loops. 1D gliding loops may become trapped, or annihilate with vacancies, but many will leave the original cascade region and reach dislocations or grain boundaries. A small proportion of them, via clustering or trapping and coalescence, is expected to form bigger defects, becoming dislocation loops visible to TEM if they grow large enough (> 2nm in diameter).
- **Route 2:** vacancy defects follow analogue fates as SIA defects in route 1, with the significant differences that: (a) the migration energy is larger (> 0.6 eV and > 1 eV for sizes above 5 vacancies) and therefore the diffusion kinetics is orders of magnitude slower; (b) the migration has a full 3D character instead of a 1D character, favouring interactions with the other defects in their immediate vicinity. Consequently, vacancy defects spontaneously form a higher density of smaller-sized defects. Nevertheless, some of them may eventually grow to form nanovoids, visible by TEM (>1.5 nm). Very few vacancies are expected to reach the grain boundary or dislocations.

Routes 1 and 2 deliberately describe radiation-induced nano-feature evolutions that take place in a relatively pure material, virtually free of solute atoms and containing essentially only C atoms or, to a lesser extent, other interstitial impurities, such as N and O, that act as traps forming complexes with vacancies [53]. The presence of solutes substantially reduces the mobility of defects (especially SIA):

- **Route 3:** Solute atoms, initially distributed in random positions in the matrix, strongly interact with migrating point defects or clusters. While the stable SIA-solute or vacancy-solute pairs that form are mobile, small SIA loops and vacancy clusters are rapidly immobilized by interacting with multiple solutes. Immobilized point-defect clusters become sinks for point defects that drag solutes, thereby providing nucleation sites for NSRC. Depending on the flux of incoming point-defects and species, and on the stability of the point-defect/solute complexes that form, the NSRC can either grow or dissolve. Since they are immobile, coalescence hardly ever occurs, thereby preventing the formation of significantly larger features: this is likely to be the reason why voids or loops are only rarely observed in RPV steels by TEM, or only at sufficiently high dose.
- **Route 4:** The NSRC formed in route 3, due to their small size (and often undefined thermodynamic phase nature) remain coherent with the lattice structure of the Fe matrix. Occasionally they may grow to such an extent that a local phase change may happen. One example could be the formation of an fcc Cu precipitate in Fe, which experimentally only takes places when a coherent Cu cluster has grown above 5 nm diameter [89].

The initial conditions coupled with the parameters that describe migration and dissociation events, together with the reaction volumes, will determine the resultant recombination, clustering, trapping and absorption rates in the simulation. If the physical parameterization is reasonable, the model will prove to be physically sound and accurately reflect the imposed operating conditions (temperature, dose-rate, dose, ...). For instance, in the range of 1-3 at% of solutes, route 3 is expected to dominate, while route 1 and route 2 will dominate in very dilute alloys.

In the model presented here, the physical parameterization is based on an extensive database of migration and binding energy values calculated by DFT. However, this database cannot cover all possible combinations of solutes and point-defects, even less all size ranges. So, a simplified description of the binding energy values to extrapolate to all cluster sizes is needed. The parameters will also be affected by the initial conditions. Parameters such as the actual initial dislocation density or concentration of carbon atoms in the matrix are often not known, or known with low precision. In these cases, the values assigned to the initial conditions are more of an assumption than an actual experimental input. Unless APT measurements are available, uncertainties affect also the concentration in the matrix of solutes that form carbides or other phases during fabrication, for example Mn. Changes on these initial concentrations of solute atoms in the matrix may have an effect on the simulation results. A full range sensitivity study should be performed for all initial condition parameters; however, this would be extremely cumbersome and out of the scope of the present work, which does not have the ambition of presenting a fully calibrated model. Instead, we aim to demonstrate that the physical assumptions of the model lead to reasonable results based on comparison with experimental data. We thus use sensible parameters to evaluate which are the dominant evolution pathways, out of all those described in section 2.1. Therefore, initial conditions such as dislocation density or C content of the matrix are here based on either known experimental data or reasonable guesses, and are the same for all steels, distinguishing only the case of model alloys, where lower C contents and dislocation densities than in steels are expected.

### 3.2 Parameterization of events related to point-defect/solute clusters

With respect to previous work [60-61,81-82], four new events must be adequately parameterized in the present OKMC model, to describe the interactions between solute atoms and point-defects, as depicted in Fig. 5.

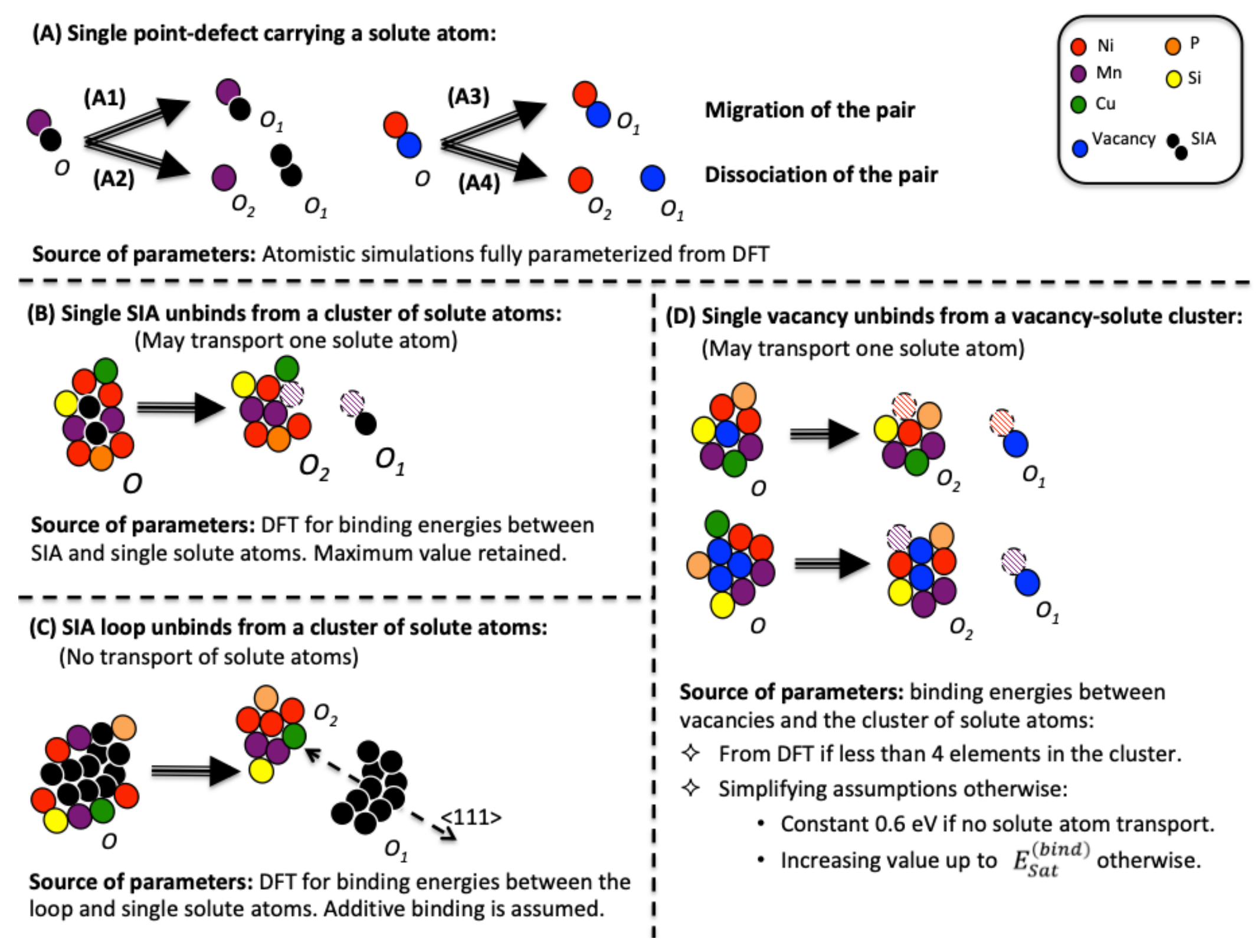

**Fig. 5 –** Definition of four events for an accurate description of the interactions between solute atoms and point-defects in our OKMC model.

Except for A1 and A3, all these events correspond to the separation of a reactant in two products, denoted as $O \rightarrow O_1 + O_2$, where by convention $O_1$ is the object that migrates away from $O_2$. In order to parameterize the events of Fig. 5 in an OKMC model, their frequency $\Gamma_e$ in Eq. 1 needs to be calculated with an activation energy expressed, in general, as:

$$E_e^{(act)} = E_{O_1}^{(mig)} + E_{O_1,O_2}^{(bind)} \tag{2}$$

Here, $e$ = A, B, C or D in Fig. 5. The first term is the migration energy of $O_1$ alone and the second one is the binding energy between $O_1$ and $O_2$. The values of $E^{(bind)}_{O_1,O_2}$ have been here calculated by DFT methods for a large number of small clusters, up to triplets. The use of DFT is necessary because no interatomic potentials [90] exist that can handle the required chemical complexity, nor are other cohesive energy models such as DFT-fitted cluster expansions [91] currently available. The shortcoming of using DFT is that the binding energy $E^{(bind)}_{O_1,O_2}$ could not be systematically evaluated for all (or many) sizes, because of the prohibitive computing time that the evaluation of the combinatorial large number of possible NSRC chemical compositions would demand. Therefore, the binding energies associated with bigger defects than those tabulated must be estimated:

$$E^{(bind)}_{O_1,O_2} = E^{(bind)}_{O_1,O_2}\Big|_{DFT} \qquad \text{if available} \tag{3}$$

$$= E^{(bind)}_{O_1,O_2}\Big|_{Est} \qquad \text{otherwise} \tag{4}$$

Here, the subscript "DFT" refers to cases where the binding energy is provided directly by DFT calculations. The values used in this work are listed in Tab. 2 for events A to C, and in the supplementary material for events D. The subscript "Est" refers to cases where the binding energy is estimated by extrapolating from the available DFT data, using simplifying assumptions that are described in Fig. 5 and in what follows.

**Events A** describe the behaviour of single point-defects when they are coupled with a Cu, Ni, Mn, Si or P solute atom. The migration of the pair towards a 1nn position, or the dissociation of the pair, is parameterized using DFT energy data derived from [48-50].

**Events B** describe the dissociation of a single SIA defect from a NSRC. This event is expected to be relatively rare. Therefore, to a first approximation, we roughly estimate the binding energy $E^{(bind)}_{O_1,O_2}$ as the maximal binding energy between the SIA and single solute atoms in the cluster, as evaluated by DFT. Note that only Mn and P atoms may be dragged away from the NSRC, because Cu, Ni and Si atoms are not energetically favourable in mixed dumbbells.

**Events C** describe how the mobility of a loop is affected by decorating solutes. Since loops cannot transport solutes on long distances while migrating [64], the only possibility left is a detachment from the whole cluster of solute atoms. This will obviously become the less likely, the more solutes decorate the loop. DFT data concerning the binding between SIA loops and individual solute atoms are provided in [52], for up to two solutes decorating the loop. The DFT data for two solutes are shown in that work to be roughly additive. We thus assume that, as long as the number of solutes is small, they will interact very little with each other, especially if located close to the edge of the loop where the interaction is the strongest. We can therefore write:

$$E^{(bind)}_{O_1,O_2}\Big|^{(event\ C)}_{Est} = \sum_{S=Cu,Ni,Mn,Si,P} E^{(bind)}_{S,n_I} \cdot n_S \tag{5}$$

Here, $E^{(bind)}_{S,n_I}$ is the maximum value (edge) provided in [52], as summarized in Tab. 2. For example, a loop decoration by a single solute atom induces a binding strength between 0.2 eV and 1 eV, while a decoration by two solute atoms induces a binding as strong as 0.4 eV to 2 eV, depending on the chemical species involved. Thus at RPV-relevant temperatures or lower a loop is effectively immobilized as soon as it is decorated by a couple of solute atoms. This is a key assumption in the model.

**Events D** describe the dissociation of a single vacancy, or a vacancy-solute pairs, from a NSRC. For simplicity, we estimated the binding energy between a NSRC and a single vacancy assuming that $E^{(bind)}_{O_1,O_2}\Big|_{Est}$ in Eq. 4 has a constant value of 0.6 eV. In contrast, the binding energy with a vacancy-solute pair was assumed to increase with the size of the NSRC, expressed in terms of the absolute number of solute atoms in the defect. Inspired from our database of DFT data, we assumed that in this case $E^{(bind)}_{O_1,O_2}\Big|_{Est}$ increases linearly with size from 0.52 eV until the NSRC contains 65 atoms. It is then assumed to have a constant, saturated value, henceforth denoted as $E^{(bind)}_{Sat}$.

**Table 2 –** List of input parameters (binding or migration energies, in eV) derived from DFT for the events A to C in Fig. 5.

| Event | Parameter | Cu | Ni | Mn | Si | P | Ref |
|---|---|---|---|---|---|---|---|
| A1 | $E^{(mig)}_{O_1}$ | 0.76 | 0.72 | 0.76 | 0.81 | 0.61 | [48, 49] |
| A2 | $E^{(mig)}_{O_1}$ | 0.63 | | | | | [81] |
| | $E^{(bind)}_{O_1,O_2}$ | 0.16 | 0.19 | 0.16 | 0.33 | 0.43 | [48, 49] |
| A3 | $E^{(mig)}_{O_1}$ | n.a. | n.a. | 0.30 | n.a. | 0.14 | [50] |
| A4 | $E^{(mig)}_{O_1}$ | 0.31 | | | | | [81] |
| | $E^{(bind)}_{O_1,O_2}$ | n.a. | n.a. | 0.55 | n.a. | 0.76 | [50] |
| B | $E^{(mig)}_{O_1}$ if no solute dragged | 0.31 | | | | | [81] |
| | $E^{(mig)}_{O_1}$ if solute dragged | n.a. | 0.30 | n.a. | 0.14 | | [50] |
| | $E^{(bind)}_{O_1,O_2}$ | 0.13 | 0.15 | 0.39 | 0.19 | 1.02 | [46] |
| C | $E^{(mig)}_{O_1}$ | 0.2 | | | | | [81] |
| | $E^{(bind)}_{S,n_I}$ | 0.4 | 0.2 | 0.37 | 0.4 | 1.06 | [52] |

## 3.3 Comparison between OKMC predictions and APT evidence

In our OKMC model a NSRC is represented by a composition vector assigned to a specific position in space. To compare with APT, this representation must be converted into an emulation of the reconstructed 3D data, as observed by the experimental technique. This is done in two steps, as illustrated in Fig. 6. As a first step, an atomic configuration needs to be created by distributing the solute atoms and the point defects in a confined region of space around the position of the cluster. To do this we deliberately assume that the solute atoms are arranged in a compact manner, filling all neighbouring positions around the central coordinates. Fe atoms are also included in the spherical cluster, in randomly chosen positions, corresponding to the number of solute-free dumbbells that were absorbed by the clusters. The second step consists of treating the data to emulate the experimental limitations associated with APT, thereby enabling a like for like comparison between simulation and experiment. During an APT experiment individual atoms are field evaporated from the sample and their positions of impact on a detector are used to determine their relative positions prior to evaporation. The process is not perfect since not all of the atoms are detected (typically between 37% and 60%, depending on the equipment) and the field evaporation is extremely complex, leading to inherent artefacts such as a degraded spatial resolution and trajectory aberrations. The detection efficiency is straightforward to emulate, namely atoms from model data are eliminated randomly to yield the appropriate detection efficiency. For simplicity, even though experiments were performed with different equipment, we assume a constant detection efficiency of 40% for all chemical species. The position uncertainty is greater in the lateral direction, with respect to the normal of the surface of the APT needle, than longitudinally [92] and can be sufficiently large to prevent resolution of extremely small clusters. To simulate this effect in the model data, a Gaussian distributed random noise (FWHM=1nm in the plane perpendicular to the needle axis, FWHM=0.1nm in the direction of the needle axis) has been applied to the atomic coordinates [92]. Finally, a simple close neighbour algorithm is applied in order to search for clusters in the simulation box. It has to be noticed that APT artefacts such as trajectory focusing are not explicitly considered here. The analysis of the simulation results, after the treatment that is described above, is less critical than the analysis of experimental APT data, because in fact precise information is available about every single atom: the randomization as described is introduced to be more consistent with the situation of the experiment. Importantly, the simulation volumes are comparable with usual APT tip volumes, i.e.~$10^5$-$10^6$ nm$^3$.

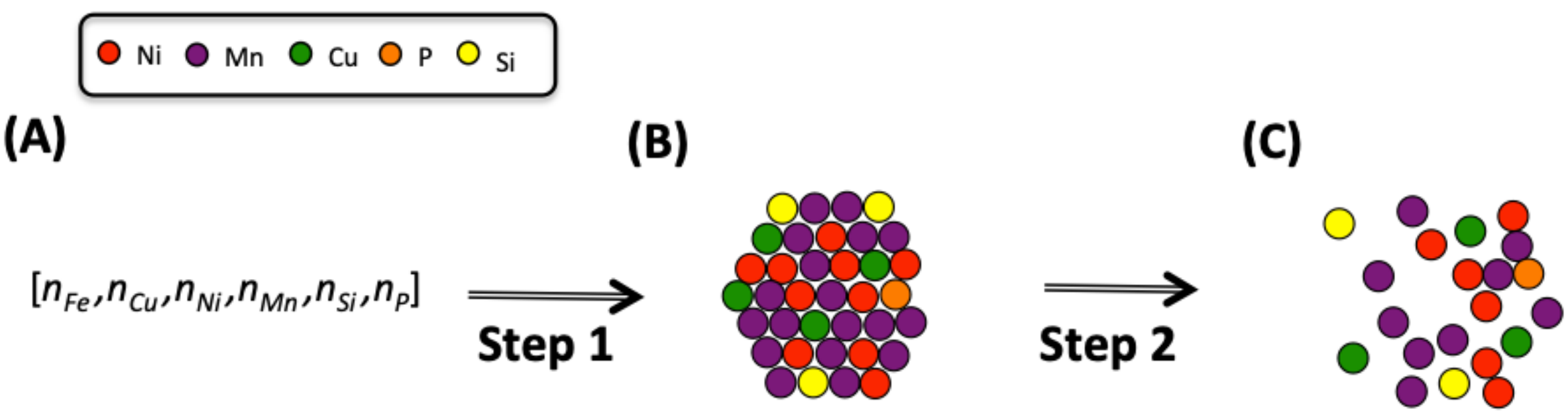

**Fig. 6** – Procedure followed for the comparison between NSRC predicted as an outcome of OKMC predictions and experimental evidence with APT. On (A) the solute clusters as described in the OKMC model (composition vector) are transformed into a compact 3D structure (B), which are later on virtually distorted as the APT measurement would do (C).

## 4 Application and results

In this section, we apply the OKMC model described in section 3, to the whole set of materials and irradiation conditions listed in Tab. 1. A few input parameters deserve preliminary discussion:

- The chemical compositions given in the table are the measured matrix values from APT and they are used as input for the model. The values differ from the nominal chemical compositions specified by the steel manufacturers, because some solutes (for example Mn) tend to form carbides. Where APT data was not available (e.g. when samples were only analysed with SANS), it was assumed that 20% of the Mn solutes were located in carbides [60-61].

- We assumed that the grain size is 75 μm and the dislocation density is $5 \cdot 10^{13}$ m$^{-2}$ for all model alloys [10, 7], which are ferritic materials, because in this case measurements exist. The ambiguous definition of the grain size in the bainitic phase and the often lacking assessment of the dislocation density before irradiation in the case of steels obliged us to assume a grain size of 1 μm and a dislocation density of $2.5 \cdot 10^{14}$ m$^{-2}$, based on only a few studies [93].
- The concentration of free C in the bulk of the bainitic phase is rarely (if ever) reported, because experimentally this measurement is not trivial. Most C is expected to precipitate in carbides, or segregate at grain boundaries and dislocations, but some remains in the matrix. In this study, we adopt similar hypotheses to those in [60-61, 81-84], namely the bulk C concentration is assumed to be 50 appm for all model alloys and 150 appm in the case of steels.

The development of a model with fully quantitative capabilities ideally requires an accurate evaluation of the impact of all the above parameters that define the initial conditions. However, the numerical values chosen here can be considered realistic and will not change over orders of magnitude. Furthermore, in this study our purpose is not to obtain the best quantitative agreement possible for each case. Instead, we aim to evaluate whether the mechanisms that have been implemented in the model, as in Fig. 1, are sufficient to explain quantitatively the experimentally observed formation of NSRC in steels under irradiation, and if so which evolution pathways dominate.

The one parameter that is mainly responsible for inaccuracies is $E_{Sat}^{(bind)}$ defined above, because of the approximation that is made by assuming that a quantity that should depend on the local atomic configuration is a constant, irrespective of the accumulated dose, i.e. irrespective of the evolution of the size and composition of the NSRC. Different values were used for this parameter, to appreciate its impact on the model predictions.
Simulations were performed in boxes with dimensions 110 $a_0$ × 120 $a_0$ × 180 $a_0$, i.e., 31.6 nm × 34.4 nm × 51.7 nm which contain 4.752 million atoms. Each simulation was in fact performed in ten independent boxes, and overall statistics were performed. In each case, the total volume that was analysed applying the procedure of comparison with APT that was described in section 3.3 was, thus, $5.62 \times 10^5$ nm$^3$, containing 47.5 million atoms. The main results are summarized in Fig. 7, where three panels compare the geometric features of the NSRC as predicted by our OKMC model with experimental evidence. The number density is affected by the numerical value of $E_{Sat}^{(bind)}$, while the average size is almost insensitive to it. Overall, one can see that the OKMC predictions fall in the experimental range, provided that the value of $E_{Sat}^{(bind)}$is in an adequate range. Remarkably, the latter appears to be small and limited to:

$$E_{Sat}^{(bind)} = 0.7 - 0.9 \text{ eV} \quad (6)$$

In specific steels the range of uncertainty is even narrower. Only a few cases exhibit a significant discrepancy with experimental evidence: (a) two materials, namely, the FeCuNiMn alloy at high dose-rate and the FZD steel at medium dose-rate show a large scatter of experimental data, which makes the comparison with the OKMC model more open to interpretation; (b) The ANP6 cases were less satisfactorily predicted by the OKMC model.

Concerning the Fe content in the NSRC, our model predicts it to be in a range between 50% and 75%. This is in reasonable agreement with APT evidence. The OKMC model also predicts that the chemical compositions of the NSRC tend to correlate with the overall chemical composition of the material. This was expected, because the driving force for their formation in our model is the transport of matrix solute atoms towards sinks, irrespective of the composition of existing clusters. The observations by APT appear to be in relatively good agreement with this result. The only clear exception is the FeCuNiMn model alloy. This is not surprising, considering that in this system strong thermodynamic driving forces are at work (Cu insolubility), which are not included in the present model. Lastly, OKMC predicts that 5% to 20% of the NSRC contain vacancies. Likewise, no more than 20% to 50% are predicted to contain SIAs, while the remaining part is predicted to be free of point-defects. From a different point of view OKMC predicts that only 5% to 15% of the vacancies are associated with solute atoms. This is in good agreement with PAS evidence [72-75].

The analysis of the details of the events that take place during the OKMC simulations helps to rationalize the different mechanisms described in Fig. 1. Strictly speaking, making a detailed balance of each transition depicted in the figure is not straightforward, because they involve a large number of actual KMC events. Overall, the model predicts the formation, from 0.01 dpa onwards, of predominantly SIA-free NSRC, as depicted in Fig. 1i: 20% of the NSRC have a vacancy content, while 60% of the NSRC are totally free of point-defects. Only the remaining 20% of the NSRC are predicted to be those depicted in Fig. 1g, i.e., small loops decorated with solutes. No big loops as depicted in Fig. 1h, were predicted. Nevertheless, depending on the case (material, irradiation condition, etc.), and the assumed value for $E_{Sat}^{(bind)}$, we estimate that 80% to 95% of all the NSRC are originally nucleated at SIA defects, i.e., following the mechanisms depicted in Fig. 1a, Fig. 1b and Fig. 1c. This can be further illustrated by analysing the sensitivity of the model predictions with respect to the events defined in section 3.2:

- Events A are the only source of solute transport. They are thus essential in the model, because no formation of NSRC would be predicted otherwise. It is worth noting that they are accurately parameterized from DFT calculated migration energies.
- Events B describe the interaction between single SIA defects and solute clusters. Our simulations reveal that these events are relatively rare, as expected in section 3.2, and they thus have a negligible influence on the model predictions.
- Events C describe the interaction between SIA loops and solute clusters. Additive binding with individual solute atoms was assumed in section 3.2. As expected, our study reveals that it is a key assumption in the model. As a test case, for the sake of demonstration, OKMC simulations were conducted dividing the $E_{O_1,O_2}^{(bind)}$ in Eq (5) by the total number of solutes decorating the loop. This is equivalent to assuming an average binding with the individual solutes, without additivity. In this case, the OKMC model predicts a number density of NSRC about an order of magnitude lower than the experimental ones.
- Events D describe the interaction between single vacancies and NSRC. The sensitivity of the model predictions with respect to the involved parameter $E_{Sat}^{(bind)}$ is illustrated in Fig. 7. The lower $E_{Sat}^{(bind)}$, the lower the number density of NSRC predicted by the OKMC model.

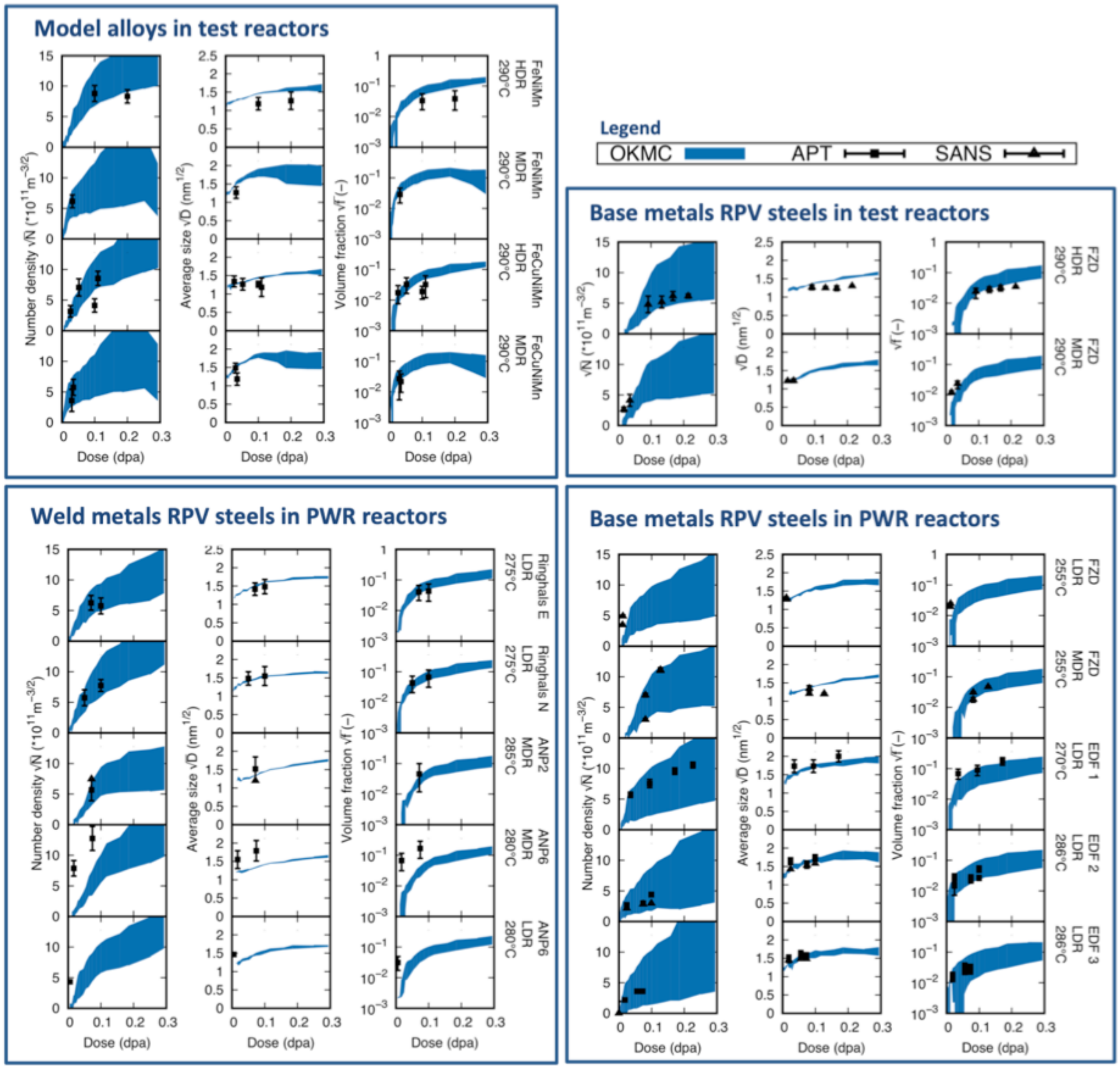

**Fig. 7 –** Summary of the prediction of our OKMC model for the formation of NSRC in all materials and irradiation conditions listed in Tab. 1. The panels summarize the geometry of the NSRC in term of number density $N$, average size $D$, and volume fraction ($f = ND^3\pi/6$) in the material. Where possible, direct comparison is done with experimental evidence from APT and/or SANS. The line thickness for OKMC predictions denotes the uncertainty with respect to the saturated binding energy $E_{Sat}^{(bind)}$.

## 5 Discussion and conclusion

Our model provides reasonable predictions of the nanostructure evolution under irradiation, as compared with post-irradiation APT and SANS examination results, in a fairly wide range of alloy compositions and irradiation conditions relevant for RPV materials under normal operation, at the temperature for which experimental data to compare with exist. This suggests that the physical mechanisms depicted in Fig. 1, on which it is based, are realistic. The key mechanisms are thus: (a) dragging of solute atoms by single point-defects; (b) heterogeneous nucleation of solute clusters on small point-defect clusters, especially SIA clusters, as revealed while analysing the results of our simulations. Strictly speaking what is depicted in Fig. 1 is not precipitation, but solute segregation at sinks. This is a consequence of the processes strongly supported by DFT calculations and atomistic simulations. The composition of the clusters is thus largely the consequence of the transport of solutes by point defects. Therefore, depending on the respective diffusivities and thus on temperature, it should bear some correlation with the alloy composition. Importantly, in this scenario the fundamental mechanisms never change over the duration of the irradiation and the NSRC develop in a continuous way. This agrees with the absence of observation of changes of embrittlement mechanisms with increasing neutron dose [94]. Notwithstanding, the model keeps the possibility open to the appearance of thermodynamically stable phases, should the local composition at sinks allow it.

Our model relies on two key assumptions. The first key assumption is the additive binding effect of individual solute atoms that decorate loops. The second key assumption is the use of a constant "saturated binding energy" (denoted as $E_{Sat}^{(bind)}$ above) between a point-defect/solute pair and a large solute or solute-vacancy cluster. This parameter is by nature local and should depend on the composition of the cluster (solutes and fraction of vacancies), but it is here assumed to be global and constant over time. Since the composition of the clusters depends on the composition of the steels, some variability of this parameter from material to material is physiological and unavoidable. This variability is indeed observed. The composition of the clusters may also be changing with radiation dose, thus the assumption of a single constant value may fail even along the same simulation in a given steel. In some cases it does happen that a wider range of variation for $E_{Sat}^{(bind)}$ is needed to match the experimental data points. However, the variability is limited to a small range of values (see Eq. 6 and discussions above) and the assumption of a single constant value fails only in a few cases, thus making it an acceptable approximation to describe the process of vacancy emission and cluster dissolution. The model is here proven valid for low-Cu operating RPV-steels, in terms of chemical composition and irradiation conditions (temperature, dose rate, dose reached and PKA energy spectrum). The application to a wider range of conditions, especially to different temperatures, would likely require a more refined parameterization in connection with both our key assumptions, i.e. additive binding for loops in Eq. 5 and constant $E_{Sat}^{(bind)}$ value in Eq. 6. Likewise, chemical compositions such as high Cu, Ni and/or Mn contents that are either known or suspected to enter in precipitation regimes could likely be correctly addressed only if the proper thermodynamics driving forces were explicitly included in the model parameterization. For all other cases, since $E_{Sat}^{(bind)}$ has a clear physical definition, it can potentially be better assessed in the future, and the approximation of a single parameter might be removed, by extending DFT calculations to larger sizes. This is currently not feasible because, in addition to computing time limitations, it entails a practical problem of data handling, as well as smart interpolation schemes, that could possibly be tackled in the future with the use of machine learning [95].

The fact that solute clusters can only dissolve after their SIA content has been removed suggests that a large proportion of clusters effectively observed in experiments should contain SIA, i.e. be associated with SIA clusters or loops. This is what our model predicts. While solute clusters will also continuously form on vacancy clusters without any SIA content, these will also be inherently less stable, because they can dissolve, thereby indeed surviving only if grown above a critical size, similarly to a 'nucleation and growth' process, and are proportionally less. These small, continuously forming and dissolving vacancy-solute clusters are unlikely to be seen by APT, due to their size, but may be identified by PAS, thanks to the chemical signatures in the environment around vacancies. Evidence of the formation of vacancy-Mn-Ni solute clusters from PAS has indeed been found [73-76]. It is unclear, however, whether these clusters can be associated with the so-called "unstable matrix damage" postulated as an additional hardening contribution at high neutron dose rate [69, 70], because it is unclear whether they will actually measurably contribute to hardening and ensuing embrittlement. MD simulations suggest that the presence of vacancies in the cluster reduces, rather than increasing, its strength [96], when interacting with edge dislocations. In contrast, when interacting with screw dislocations the formation of helical turns may result in a strong hindrance to dislocation motion [97]. It is difficult to anticipate which effect will be dominant. In contrast, it should be considered as demonstrated that dislocation loops that are decorated by solutes exhibit higher strength as obstacles than undecorated loops [98], thus stable NSRC certainly do cause hardening.

In a recent work, Ke *et al.* [43] proposed a cluster dynamics model for RPV steels based on the underlying idea that irradiation mainly accelerates the kinetics of phase separation due to vacancy supersaturation. They thus inherently assumed that the NSRC are precipitates of thermodynamically stable phases, following the predictions of Thermo-Calc [67]. As a correcting factor, the model of Ke *et al.* also included an extra term generically associated with "radiation-induced heterogeneous nucleation of precipitates on point defect clusters produced in displacement cascades". Interestingly, the fitting of the model to experimental data revealed that the latter term is dominant in most cases over the thermodynamic component, in some cases by orders of magnitude. This result is thus in good agreement with the result of the present study. In spite of putting emphasis on two different aspects (thermodynamic driving forces or solute transport and heterogeneous nucleation by segregation), both approaches therefore strongly suggest that SIA defects play a dominant role in NSRC formation and evolution under irradiation. However, a number of problems still need to be addressed for the model to be reliably generalized to concentrated alloys such as high-Cr ferritic-martensitic steels or austenitic steels. Moreover, segregation processes at extended defects cannot be treated in the current framework of effective sink description, while high doses (tens of dpa) remain challenging because of the still high computing cost of the simulations. What is possible in terms of model improvement in the short term are sensitivity studies and application to simulate post irradiation annealing to assess the thermal stability of the nanostructural features. These studies will be instrumental for the application to a wider range of irradiation or annealing temperatures and the comparison with experimental data from somewhat larger accumulated doses, as final proof of validity for the characteristic energy description in the entire parameterization of the model. Notwithstanding, when coupled to existing physical models for the assessment of radiation hardening [35-36], the present model is expected to provide a powerful tool for the assessment of radiation-induced hardening in irradiated RPV steels under a variety of irradiation conditions. This includes conditions of relevance for long term operation (for which experimental data are limited), the comparison between ion or neutron irradiation, and the assessment of flux and temperature effects.

## Acknowledgements

This work has received partial funding from the Euratom research and training programme 2014-2018 under Grant Agreement No. 661913 (SOTERIA project: Safe long-term operation of light water reactors based on improved understanding of radiation effects in nuclear structural materials) and Grant Agreement No. 249360 (LONGLIFE project: Treatment of long term irradiation embrittlement effects in RPV safety assessment). The views and opinions expressed herein do not necessarily reflect those of the European Commission.

## References

[1] P. Petrequin, A review of formulas for predicting irradiation embrittlement of reactors vessel materials, *AMES Report* No. 6, EUR 16455, European Commission, Luxembourg (November 1996).

[2] C. English, J. Hyde, Radiation damage of reactor pressure vessel steels, in: R.J.M. Konings (Editor-in-chief), *Comprehensive Nuclear Materials* Vol. 4: Radiation effects in structural and functional materials for fission and fusion, Elsevier Ltd., 2012, pp. 151-180.

[3] W.L. Server, R.K. Nanstad, G.R. Odette, Use of reactor pressure vessel surveillance materials for extended life evaluations using power and test reactor irradiations. Third International Conference on Nuclear Power Plant Life Management (PLiM) for Long Term Operations (LTO), *International Atomic Energy Agency (IAEA),* 2012. https://inis.iaea.org/collection/NCLCollectionStore/_Public/43/070/43070853.pdf

[4] A. Ulbricht, E. Altstadt, F. Bergner, H.-W. Viehrig, U. Keiderling, Small-angle neutron scattering investigation of as-irradiated, annealed and reirradiated reactor pressure vessel weld material of decommissioned reactor, *J. Nucl. Mat.* 416 (2011) 111-116.

[5] R. Chaouadi, R. Gérard, Neutron flux and annealing effects on irradiation hardening of RPV materials, *J. Nucl. Mater.* 418 (2011) 137-132.

[6] P. Efsing, J. Roudén, P. Nilsson, Flux effects on radiation induced aging behaviour of low alloy steel weld material with high nickel and manganese content, in: M. Kirk, E. Lucon (Eds.), Effects of Radiation on Nuclear Materials: 26th Volume, ASTM International, West Conshohocken, 2014, pp. 119-134. https://doi.org/10.1520/STP157220130112.

[7] L. Malerba, E. van Walle, C. Domain, S. Jumel, J.-C. van Duysen, State of advancement of the International REVE Project: Computational Modelling of Irradiation-Induced Hardening in Reactor Pressure Vessel Steels and Relevant Experimental Validation Programme. Proceedings 10th Intl. Conference on Nuclear Engineering (ICONE-10), Arlington, VA, USA, April 14-18, 2002, *The American Society of Mechanical Engineers*, 2002, pp. 267-274.

[8] J.-P. Massoud, S. Bugat, B. Marini, D. Lidbury, S. van Dyck, PERFECT – Prediction of Irradiation Damage Effects on Reactor Components: A summary, *J. Nucl. Mater.* 406 (210) 2-6.

[9] A. Al-Mazouzi, A. Alamo, D. Lidbury, D. Moinereau, S. van Dyck, PERFORM 60: Prediction of the effects of radiation for reactor pressure vessel and in-core materials using multi-scale modelling – 60 years foreseen plant lifetime, *Nucl. Eng. Design* 241 (2011) 3403-3415.
[10] E. Meslin, B. Radiguet, P. Pareige, A. Barbu, Kinetic of solute clustering in neutron irradiated ferritic model alloys and a French pressure vessel steel investigated by atom probe tomography, *J. Nucl. Mater.* 399 (2010) 137-145.
[11] P. Auger, P. Pareige, M. Akamatsu, D. Blavette, APFIM investigation of clustering in neutron-irradiated Fe-Cu alloys and pressure vessel steels. *J. Nucl. Mater.* 225 (1995) 225-230.
[12] P. Auger, P. Pareige, S. Welzel, J.-C. van Duysen, Synthesis of atom probe experiments on irradiation-induced solute segregation in French ferritic pressure vessel steels. *J. Nucl. Mater.* 280 (2000) 331-344.
[13] M.K. Miller, K.F. Russell, J. Kocik, E. Keilova, Embrittlement of low copper VVER 440 surveillance samples neutron-irradiated to high fluences. *J. Nucl. Mater.* 282 (2000) 83-88.
[14] K. Fukuya, K. Ohno, H. Nakata, S. Dumbill, J.M. Hyde, Microstructural evolution in medium copper low alloy steels irradiated in a pressurized water reactor and a material test reactor. *J. Nucl. Mater.* 312 (2003) 163-173.
[15] P. Pareige, R.E. Stoller, K.F. Russell, M.K. Miller, Atom probe characterization of the microstructure of nuclear pressure vessel surveillance materials after neutron irradiation and after annealing treatments, *J. Nucl. Mater.* 249 (1997) 165-164.
[16] H. Huang, B. Radiguet, P. Todeschini, G. Chas, P. Pareige, Atom Probe Tomography characterization of the microstructural evolution of a low coper reactor pressure vessel steel under neutron irradiation. *Mater. Res. Soc. Symp. Proc.* 1264 (2010) BB05-18.
[17] M.K. Miller, A.A. Chernobaeva, Y.I. Shtrombakh, K.F. Russell, R.K. Nanstad, D.Y. Erak, O.O. Zabusov, Evolution of the nanostructure of VVER-1000 RPV materials under neutron irradiation and post irradiation annealing, *J. Nucl. Mater.* 385 (2009) 615–622.
[18] M.K. Miller, K.F. Russell, Embrittlement of RPV steels: An atom probe tomography perspective. *J. Nucl. Mater.* 371 (2007) 145-160.
[19] B.A. Gurovich, E.A. Kuleshova, Ya.I. Shtrombakh, D.Yu. Erak, A.A. Chernobaeva, O.O. Zabusov, Fine structure behaviour of VVER-1000 RPV materials under irradiation, *J. Nucl. Mater.* 389 (2009) 490-496.
[20] M.K. Miller, K.F. Russell, M.A. Sokolov, R.K. Nanstad, Atom probe tomography characterization of radiation-sensitive KS-01 weld. *J. Nucl. Mater.* 320 (2003) 177-183.
[21] E. Meslin, M. Lambrecht, M. Hernández-Mayoral, F. Bergner, L. Malerba, P. Pareige, B. Radiguet, A. Barbu, D. Gómez-Briceño, A. Ulbricht, A. Almazouzi, Characterization of neutron-irradiated ferritic model alloys and a RPV steel from combined APT, SANS, TEM and PAS analyses, *J. Nucl. Mater.* 406 (2010) 73-83.
[22] M.K. Miller, K.A. Powers, R.K. Nanstad, P. Efsing, Atom probe tomography characterizations of high nickel, low copper surveillance RPV welds irradiated to high fluences. *J. Nucl. Mater.* 437 (2013) 107-115.

[23] P.D. Edmondson, M.K. Miller, K.A. Powers, R.K. Nanstad, Atom probe tomography characterization of neutron irradiated surveillance samples from the R. E. Ginna reactor pressure vessel, *J. Nucl. Mater.* 470 (2016) 147-154.
[24] R.G. Carter, N. Soneda, K. Dohi, J.M. Hyde, C.A. English, W.L. Server, Microstructural characterization of irradiation-induced Cu-enriched clusters in reactor pressure vessel steels, *J. Nucl. Mater.* 298 (2001) 211-224.
[25] P.A. Beaven, F. Frisius, R. Kampmann, R. Wagner, J.R. Hawthorne, SANS investigation of irradiated A533-B steels doped with phosphorus, in: L.E. Steele (Ed.), Radiation embrittlement of nuclear reactor pressure vessel steels: An international review, ASTM STP 1011, ASTM International, West Conshohocken, 1989, pp. 243-256. https://doi.org/10.1520/STP10399S
[26] M.K. Miller, B.D. Wirth, G.R. Odette, Precipitation in neutron-irradiated Fe–Cu and Fe–Cu–Mn model alloys: a comparison of APT and SANS data, *Mater. Sci. Eng.* A353 (2003) 133-139.
[27] F. Bergner, M. Lambrecht, A. Ulbricht, A. Almazouzi, Comparative small-angle neutron scattering study of neutron-irradiated Fe, Fe-based alloys and a pressure vessel steel, *J. Nucl. Mater.* 399 (2010) 129-136.
[28] A. Wagner, A. Ulbricht, F. Bergner, E. Altstadt, Influence of the copper impurity level on the irradiation response of reactor pressure vessel steels investigated by SANS, *Nucl. Instr. Meth. Phys. Res. B* 280 (2012) 98-102.
[29] F. Bergner, C. Pareige, M. Hernández-Mayoral, L. Malerba, C. Heintze, Application of a three-feature dispersed-barrier hardening model to neutron-irradiated Fe–Cr model alloys, *J. Nucl. Mater.* 448 (2014) 96-102.
[30] M. Lambrecht, E. Meslin, L. Malerba, M. Hernández-Mayoral, F. Bergner, P. Pareige, B. Radiguet, A. Almazouzi, On the correlation between irradiation-induced microstructural features and the hardening of reactor pressure vessel steels, *J. Nucl. Mater.* 406 (2010) 84-89.
[31] Z. Lu, R. Faulkner, R. Jones, P. Flewitt, Radiation- and thermally-induced phosphorus inter-granular segregation in pressure vessel steels, *J. ASTM International*, 2 (2005) 1-15. https://doi.org/10.1520/JAI12387
[32] C. Pareige, V. Kuksenko, P. Pareige, Behaviour of P, Si, Ni impurities and Cr in self ion irradiated Fe–Cr alloys – Comparison to neutron irradiation, *J. Nucl. Mater.* 456 (2015) 471-476.
[33] P.D. Edmondson, C.M. Parish, R.K. Nanstad, Using complimentary microscopy methods to examine Ni-Mn-Si precipitates in highly-irradiated reactor pressure vessel steels, *Acta Mater.* 134 (2017) 31-39.
[34] N. Soneda, K. Dohi, A. Nomoto, K. Nishida, S. Ishino, Embrittlement correlation method for the Japanese reactor pressure vessel materials, *J. ASTM International* 7 (2010) 1-10. https://doi.org/10.1520/JAI102127
[35] G. Monnet, Multiscale modeling of precipitation hardening: Application to the Fe–Cr alloys, *Acta Mater.* 95 (2015) 302-311.
[36] G. Monnet, Multiscale modeling of irradiation hardening: Application to important nuclear materials, *J. Nucl. Mater.* 508 (2018) 609-627.
[37] R.S. Averback, T. Diaz de la Rubia, Displacement damage in irradiated metals and semiconductors, *Solid State Physics* 51 (1997) 281-402.
[38] R.E. Stoller, A.F. Calder, Statistical analysis of a library of molecular dynamics cascade simulations in iron at 100 K, *J. Nucl. Mater.* 283 (2000) 746-752.
[39] L. Malerba, Molecular dynamics simulation of displacement cascades in a-Fe: A critical review, *J. Nucl. Mater.* 351 (2006) 28-38.

[40] G.R. Odette, B.D. Wirth, A computational microscopy study of nanostructural irradiated pressure vessel steels, *J. Nucl. Mater.* 251 (1997) 157-171.

[41] L.C. Liu, G.R. Odette, B.D. Wirth, G.E. Lucas, A lattice Monte Carlo simulation of nanophase compositions and structures in irradiated pressure vessel Fe-Cu-Ni-Mn-Si steels, *Mater. Sci. Eng. A* 238 (1997) 202-209.

[42] P.B. Wells, T. Yamamoto, B. Miller, T. Milot, J. Cole, Y. Wu, G.R. Odette, Evolution of manganese–nickel–silicon-dominated phases in highly irradiated reactor pressure vessel steels, *Acta Mater.* 80 (2014) 205-219.

[43] H. Ke, P. Wells, P.D. Edmondson, N. Almirall, L. Barnard, G.R. Odette, D. Morgan, Thermodynamic and kinetic modeling of Mn-Ni-Si precipitates in low-Cu reactor pressure vessel steels, *Acta Mater*. 138 (2017) 10-26.

[44] M. Mamivand, P. Wells, H. Ke, S. Shu, G.R. Odette, D. Morgan, CuMnNiSi precipitate evolution in irradiated reactor pressure vessel steels: Integrated Cluster Dynamics and experiments, *Acta Mater.* 180 (2019) 199-217.

[45] N. Almirall, P.B. Wells, T. Yamamoto, K. Wilford, T. Williams, N. Riddle, G.R. Odette, Precipitation and hardening in irradiated low alloy steels with a wide range of Ni and Mn compositions, *Acta Mater.* 179 (2019) 119-128.

[46] R. Ngayam-Happy, C.S. Becquart, C. Domain, First principle-based AKMC modeling of the formation and medium-term evolution of point defect and solute-rich clusters in a neutron irradiated complex Fe-CuNiMnSiP alloy representative of reactor pressure vessel steels, J. Nucl. Mater. 440 (2013) 143-152.

[47] R. Ngayam-Happy, C.S. Becquart, C. Domain, L. Malerba, Formation and evolution of MnNi clusters in neutron irradiated dilute Fe alloys by a first principle-based AKMC method, *J. Nucl. Mater.* 426 (2012) 198-207.

[48] L. Messina, M. Nastar, T. Garnier, C. Domain, P. Olsson, Exact *ab initio* transport coefficients in bcc Fe−X (X=Cr, Cu, Mn, Ni, P, Si) dilute alloys, *Phys. Rev. B* 90 (2014) 104203.

[49] L. Messina, M. Nastar, N. Sandberg, P. Olsson, Systematic electronic-structure investigation of substitutional impurity diffusion and flux coupling in bcc iron, *Phys. Rev. B* 93 (2016) 184302.

[50] L. Messina, T. Schuler, M. Nastar, M.-C. Marinica, P. Olsson, Solute diffusion by self-interstitial defects and radiation-induced segregation in ferritic Fe–X (X=Cr, Cu, Mn, Ni, P, Si) dilute alloys, *Acta Mater*. 191 (2020) 166.

[51] C.S. Becquart, R. Ngayam-Happy, P. Olsson, C. Domain, A DFT study of the stability of SIAs and small SIA clusters in the vicinity of solute atoms in Fe, *J. Nucl. Mater.* 500 (2018) 92.

[52] C. Domain, C.S. Becquart, Solute -<111> interstitial loop interaction in α-Fe: A DFT study, *J. Nucl. Mater.* 499 (2018) 582-594.

[53] C. Barouh, T. Schuler, C.-C. Fu, M. Nastar, Interaction between vacancies and interstitial solutes (C, N, and O) in *α*-Fe: From electronic structure to thermodynamics, *Phys. Rev. B* 90 (2014) 054112.

[54] A. Vehanen, P. Hautojärvi, J. Johansson, J. Yli-Kauppila, P. Moser, Vacancies and carbon impurities in α-iron: Electron irradiation, *Phys. Rev. B* 25 (1982) 762-780.

[55] B. Minov, M. Lambrecht, D. Terentyev, C. Domain, M.J. Konstantinović, Structure of nanoscale copper precipitates in neutron-irradiated Fe-Cu-C alloys, *Phys. Rev. B* 85 (2012) 024202.

[56] D. Terentyev, A. Bakaev, E.E. Zhurkin, Effect of carbon decoration on the absorption of < 100> dislocation loops by dislocations in iron, *J. Phys.: Condens. Matter*. 26(2014) 165402.

[57] N. Anento, A. Serra, Carbon–vacancy complexes as traps for self-interstitial clusters in Fe–C alloys, *J. Nucl. Mater.* 440 (2013) 236-242.

[58] G. Bonny, D. Terentyev, E.E. Zhurkin, L. Malerba, Monte Carlo study of decorated dislocation loops in FeNiMnCu model alloys, *J. Nucl. Mater.* 452 (2014) 486-492.

[59] M.I. Pascuet, E. Martínez, G. Monnet, L. Malerba, Solute effects on edge dislocation pinning in complex alpha-Fe alloys, *J. Nucl. Mater.* 494 (2017) 311-321.

[60] M. Chiapetto, L. Malerba, C.S. Becquart, Nanostructure evolution under irradiation in FeMnNi alloys: A "grey alloy" object kinetic Monte Carlo model, *J. Nucl. Mater.* 462 (2015) 91-99.

[61] M. Chiapetto, C.S. Becquart, C. Domain, L. Malerba, Nanostructure evolution under irradiation of Fe(C)MnNi model alloys for reactor pressure vessel steels, *Nucl. Instrum. Methods Phys. Res. B* 352 (2015) 56–60.

[62] B.N. Singh, S.I. Golubov, H. Trinkaus, A. Serra, Yu.N. Osetsky, A.V. Barashev, Aspects of microstructure evolution and cascade damage conditions, *J. Nucl. Mater.* 251 (1997) 107-122.

[63] D. Terentyev, L. Malerba, M. Hou, Dimensionality of interstitial cluster motion in bcc-Fe, *Phys. Rev. B* 75 (2007) 104108.

[64] D. Hull, D.J. Bacon, *Introduction to dislocations*, Elsevier, Amsterdam, 2011, Ch. 3.

[65] L.T. Belkacemi, E. Meslin, B. Décamps, B. Radiguet, J. Henry, Radiation-induced bcc-fcc phase transformation in a Fe-3%Ni alloy, *Acta Mater*. 161 (2018) 61-72.

[66] A.H. Delandar, O.I. Gorbatov, M. Selleby, Yu.N. Gornostyrev, P.A. Korzhavyi, Ab-initio based search for late blooming phase compositions in iron alloys, *J. Nucl. Mater.* 509 (2018) 225-236; A. Jacob, C. Domain, G. Adjanor, P. Todeschini, E. Povoden-Karadeniz, Thermodynamic modeling of G-phase and assessment of phase stabilities in reactor pressure vessel steels and cast duplex stainless steels, *J. Nucl. Mater.* 533 (2020) 152091.

[67] J.O. Andersson, T. Helander, L. Höglund, P.F. Shi, B. Sundman, Thermo-Calc and DICTRA, Computational tools for materials science, *Calphad* 26 (2002) 273-312.

[68] Y. Abe, S. Jitsukawa, Lowest energy structures of self-interstitial atom clusters in iron from a combination of Langevin molecular dynamics and the basin-hopping technique, *Phil. Mag.* 89 (2009) 375-388.

[69] G.R. Odette, R.K. Nanstad, Predictive reactor pressure vessel steel irradiation embrittlement models: issues and opportunities, *JOM* 61 (2009) 17-23.

[70] M. Burke, R. Stofanak, J. Hyde, C. English, W. Server, Microstructural aspects of irradiation damage in A508 grade 4N forging steel: composition and flux effects, *J. ASTM International* 5 (2004) 1-14.

[71] E. Altstadt, E. Keim, H. Hein, M. Serrano, F. Bergner, H.-W. Viehrig, A. Ballesteros, R. Chaouadi, K. Wilford, FP7 Project LONGLIFE: Overview of results and implications, *Nucl. Eng. Des.* 278 (2014) 753-757.

[72] M. Valo, R. Krause, K. Saarinen, P. Hautojarvi, J.R. Hawthorne, Irradiation response and annealing behaviour of pressure vessel model steels and iron ternary alloys measured with positron techniques, in: R. Stoller, A. Kumar, D. Gelles (Eds.), Effects of Radiation on Materials: 15th International Symposium, ASTM International, West Coshohocken, 1992, pp. 172-185. https://doi.org/10.1520/STP17868S

[73] Y. Nagai, Z. Tang, M. Hassegawa, T. Kanai, M. Saneyasu, Irradiation-induced Cu aggregations in Fe: An origin of embrittlement of reactor pressure vessel steels, *Phys. Rev. B* 63 (2001) 134110.
[74] S.C. Glade, B.D. Wirth, G.R. Odette, P. Asoka-Kumar, Positron annihilation spectroscopy and small angle neutron scattering characterization of nanostructural features in high-nickel model reactor pressure vessel steels, *J. Nucl. Mater.* 351 (2006) 197-208.
[75] M.J. Konstantinović, G. Bonny, Thermal stability and the structure of vacancy–solute clusters in iron alloys, *Acta Mater.* 85 (2015) 107-111.
[76] M.J. Konstantinovic, I. Uytdenhouwen, G. Bonny, N. Castin, L. Malerba, P. Efsing, Radiation induced solute clustering in high-Ni reactor pressure vessel steel, *Acta Mater.* 179 (2019) 183-189.
[77] F. Gillemot, M. Horváth, A. Horváth, B. Kovács, B. Radiguet, S. Cammelli, P. Pareige, M.H. Mayoral, A. Ulbricht, N. Kresz, F. Oszwald, G. Török, Microstructural changes in highly irradiated 15Kh2MFA steel, in: M. Kirk, E. Lucon (Eds.), Effects of Radiation on Nuclear Materials: 26th Volume, ASTM STP 1572, ASTM International, West Coshohocken, 2014, pp. 45-56. https://doi.org/10.1520/STP157220130098
[78] E.A. Marquis, Atom probe tomography applied to the analysis of irradiated microstructures. *J. Mater Res.* 30 (2015) 1222-1230.
[79] E. Meslin, B. Radiguet, M. Loyer-Prost, Radiation-induced precipitation in a ferritic model alloy: An experimental and theoretical study, *Acta Mater.* 61 (2013) 6246-6254.
[80] E. Meslin, B. Radiguet, P. Pareige, C. Toffolon, A. Barbu, Irradiation-induced solute clustering in a low nickel FeMnNi ferritic alloy, *Experimental Mechanics* 51 (2011) 1453-1458.
[81] V. Jansson, L. Malerba, Simulation of the nanostructure evolution under irradiation in Fe–C alloys, *J. Nucl. Mater.* 443 (2013) 274-285.
[82] C. Domain, C.S. Becquart, Object KineticMonte Carlo (OKMC): A Coarse-Grained Approach to Radiation Damage", Springer Nature Switzerland AG 2018, W. Andreoni, S. Yip (eds.), *Handbook of Materials Modeling*, https://doi.org/10.1007/978-3-319-42913-7_101-1.
[83] K.A. Fichthorn, W.H. Weinberg, Theoretical foundations of dynamical Monte Carlo simulations, *J. Chem. Phys.* 95 (1991) 1090-1096.
[84] M. Chiapetto, L. Malerba, C.S. Becquart, Effect of Cr content on the nanostructural evolution of irradiated ferritic/martensitic alloys: An object kinetic Monte Carlo model. *J. Nucl. Mater.* 465 (2015) 326-336.
[85] V. Jansson, L. Malerba, A. De Backer, C.S. Becquart, C. Domain, Sink strength calculations of dislocations and loops using OKMC, *J. Nucl. Mater.* 442 (2013) 218.
[86] ASTM E693-17, Standard Practice for Characterizing Neutron Exposures in Iron and Low Alloy Steels in Terms of Displacements Per Atom, ASTM International, West Conshohocken, PA 2017. https://doi.org/10.1520/E0693-17
[87] M.J. Norgett, M.T. Robinson, I.M. Torrens, A proposed method of calculating displacement dose rates, *Nucl. Eng. Des.* 33 (1975) 50-54.
[88] C. Domain, C.S. Becquart, L. Malerba, Simulation of radiation damage in Fe alloys: an object kinetic Monte Carlo approach, *J. Nucl. Mater.* 335 (2004) 121.
[89] P.J. Othen, M.L. Jenkins, G.D.W. Smith, High-resolution electron microscopy studies of the structure of Cu precipitates in α-Fe, *Philos. Mag. A* 70 (1994) 1-24.

[90] G. Bonny, D. Terentyev, A. Bakaev, E.E. Zhurkin, M. Hou, D. Van Neck, L. Malerba, On the thermal stability of late blooming phases in reactor pressure vessel steels: An atomistic study, *J. Nucl. Mater.* 442 (2013) 282-291.
[91] G. Bonny, C. Domain, N. Castin, P. Olsson, L. Malerba, The impact of alloying elements on the precipitation stability and kinetics in iron based alloys: An atomistic study, *Comput. Mater. Sci.* 161 (2019) 309-320.
[92] J.M. Hyde, G. DaCosta, C. Hatzoglou, H. Weekes, B. Radiguet, P.D. Styman, F. Vurpillot, C. Pareige, A. Etienne, G. Bonny, N. Castin, L. Malerba, P. Pareige, Analysis of radiation damage in light water reactors: Comparison of cluster analysis methods for the analysis of atom probe data, *Microsc. Microanal.* 23 (2017) 366-375.
[93] L. Zhang, B. Radiguet, P. Todeschini, C. Domain, Y. Shen, P. Pareige, Investigation of solute segregation behavior using a correlative EBSD/TKD/APT methodology in a 16MND5 weld, J. Nucl. Mater. 523 (2019) 434-443
[94] Ortner, S. et al., private communication within the LONGLIFE project **(**2014)**.**
[95] N. Castin, M.I. Pascuet, L. Messina, C. Domain, P. Olsson, R.C. Pasianot, L. Malerba, Advanced atomistic models for radiation damage in Fe-based alloys: Contributions and future perspectives from artificial neural networks, *Comput. Mater. Sci.* 148 (2018) 116-130.
[96] D. Terentyev, L. Malerba, G. Bonny, A.T. Al-Motasem, M. Posselt, Interaction of an edge dislocation with Cu-Ni-Vacancy clusters in bcc iron, *J. Nucl. Mater.* 419 (2011) 134-139.
[97] D. Terentyev, L. Malerba, Interaction of a screw dislocation with Cu-precipitates, nanovoids and Cu–vacancy clusters in BCC iron, Journal of Nuclear Materials 421 (2012) 32–38.
[98] D. Terentyev, X. He, G. Bonny, A. Bakaev, E. Zhurkin, L. Malerba, Hardening due to dislocation loop damage in RPV model alloys: Role of Mn segregation, *J. Nucl. Mater.* 457 (2015) 173–181